\documentclass[prx,twocolumn,amsmath,amssymb,floatfix,footinbib,bibnotes]{revtex4-1}

\pdfoutput=1

\usepackage[colorlinks=true,citecolor=blue,linkcolor=magenta]{hyperref}
\usepackage{amsmath}
\usepackage{graphicx}
\usepackage{dsfont}
\usepackage{changepage}
\usepackage{appendix}
\usepackage[T1]{fontenc}
\usepackage[latin1]{inputenc}
\usepackage{lipsum}
\usepackage{float}

\newcommand{\be}{\begin{equation}}
\newcommand{\ee}{\end{equation}}

\def \ua{{\uparrow}}
\def \da{{\downarrow}}
\def \be{\begin{equation}}
\def \ee{\end{equation}}
\def \ba{\begin{array}}
\def \ea{\end{array}}
\def \bea{\begin{eqnarray}}
\def \eea{\end{eqnarray}}
\def \nn{\nonumber}
\def \half{{1\over 2}}

\def \e{{\epsilon}}
\def \lam{{\lambda}}

\def \a{{\alpha}}
\def \t{{\theta}}

\def \D{{\Delta}}
\def \d{{\delta}}

\def \s{{\sigma}}
\def \f{{\varphi}}

\def \e{{\epsilon}}

\def \G{{\Gamma}}

\def \nd{{^{\vphantom{\dagger}}}}
\def \yd{^\dagger}
\def \av#1{{\langle#1\rangle}}
\def \ket#1{{\,|\,#1\,\rangle\,}}
\def \bra#1{{\,\langle\,#1\,|\,}}

\def \beas{\begin{eqnarray*}}
\def \eeas{\end{eqnarray*}}

\def \half{{\frac{1}{2}}}

\def \mrm{\mathrm}

\def \bs{\boldsymbol}
\def \mc{\mathcal}

\newcounter{indice}

\begin{document}
\title{Competition Between Kondo Screening and Magnetism at the LaAlO$_3$/SrTiO$_3$ Interface}

\author{Jonathan Ruhman, Arjun Joshua, Shahal Ilani and Ehud Altman \\
{\small \em Department of Condensed Matter Physics, Weizmann Institute of Science, Rehovot 76100, Israel}}
\begin{abstract}
We present a theory of  magnetic and magneto-transport phenomena at LaAlO$_3$/SrTiO$_3$ interfaces, which as a central ingredient includes coupling between the conduction bands and local magnetic moments originating from charge traps at the interface. Tuning  the itinerant electron density in the model drives transitions between a heavy Fermi liquid phase with screened moments and various magnetic states. The dependence of the magnetic phenomena on the electron density or gate voltage stems from competing magnetic interactions between the local moments and the different conduction bands. At low densities only the lowest conduction band, composed of the $d_{xy}$ orbitals of Ti, is occupied. Its antiferromagnetic interaction with the local moments leads to screening of the moments at a Kondo scale that increases with density. However, above a critical density, measured in experiments to be $n_c\approx 1.7\times 10^{13} cm^{-2}$,  the $d_{xz}$ and $d_{yz}$ bands begin to populate. Their ferromagnetic interaction with the local moments competes with the antiferromagnetic interaction of the $d_{xy}$ band leading to eventual reduction of the Kondo scale with density. We explain the distinct magneto transport regimes seen in experiments as manifestations of the magnetic phase diagram computed from the model. We present new data showing a relation between the anomalous Hall effect and the resistivity in the system. The data strongly suggests that the concentration of local magnetic moments affecting the transport in the system is much lower than the carrier density, in accord with the theoretical model.
\end{abstract}
\maketitle
\section{Introduction}

The conducting interface between two insulating non magnetic oxides, LaAlO$_3$ (LAO) and SrTiO$_3$ (STO)\cite{Ohtomo2004a} displays a fascinating range of emergent electronic phenomena~\cite{Hwang2012} including  superconductivity \cite{Reyren2007b,Caviglia2008a} signatures of magnetism~\cite{Brinkman2007a,Ariando2011,Bert2011,Li2011:coexistence,Lee2013} and unconventional transport properties ~\cite{Thiel2006a,BenShalom2010,Joshua2012,Joshua2013}.
One of the features that make this system particularly interesting as a new platform for investigating both basic science and technological applications is the high tunability of the electronic properties.
Many of the properties including conductivity~\cite{Thiel2006a}, superconductivity~\cite{Caviglia2008a}, spin-orbit coupling~\cite{Caviglia2010a,BenShalom2010}, orbital content~\cite{Joshua2012} as well as signatures of magnetism~\cite{Joshua2013,Bi:arXiv:2013} can be varied using a back gate, which controls the electron density at the interface.

Recent measurements using a scanning magnetic force microscope on a sample with carrier density tuned by a back gate, found ferromagnetic domains to be present at the interface at very low densities~\cite{Bi:arXiv:2013}. The magnetic domains disappear when the density is increased using the gate. Transport experiments performed with an in-plane magnetic field indicate that another (meta-magnetic) transition occurs at even higher density of conduction electrons~\cite{BenShalom2010,Joshua2013}. Above a density dependent critical field the resistance drops sharply, an anomalous Hall effect appears and the system develops a strong in-plane anisotropy of the magnetoresistance. An important goal is to understand the origin of the different transport and magnetic properties and how they depend on the density and magnetic field as well as on the microscopic building blocks of the system.

\begin{figure}
\centering
\includegraphics[width=8cm,height=6.5cm]{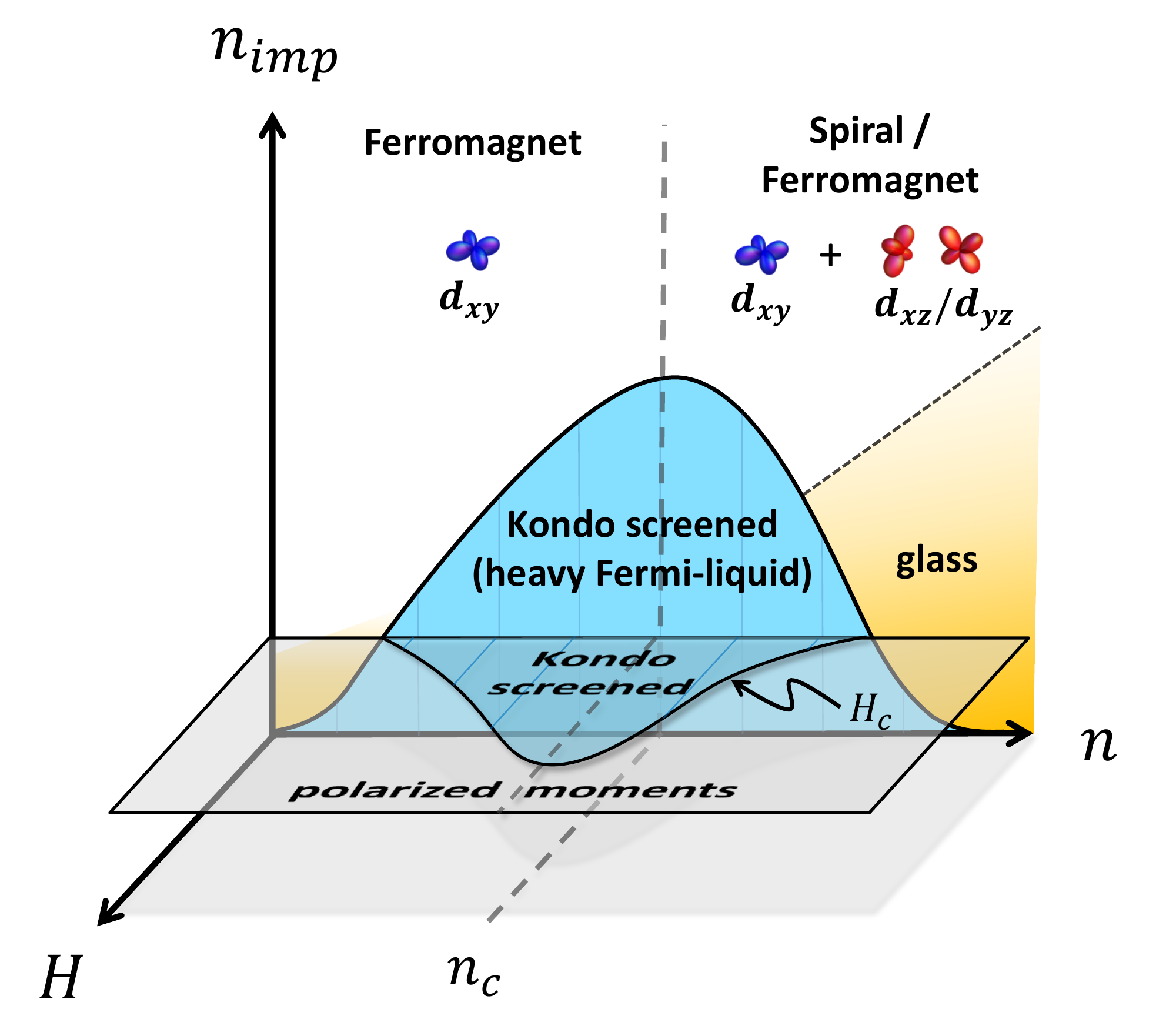}
\caption{Schematic phase diagram of the magnetic phases in the LAO/STO interface in the space of itinerant electron density ($n$), local magnetic moment density ($n_{\mrm{imp}}$) and magnetic field ($H$). The dashed line markers the critical density $n_c$ above which $d_{xz}/d_{yz}$ itinerant electrons are present in the system.  }\label{fig:pd:overview}
\end{figure}

In this paper we present a theory,  which explains the different regimes observed in experiments~\cite{BenShalom2010,Joshua2013,Bi:arXiv:2013} as manifestations of distinct electronic phases at the interface and clarifies the dependence of the phase boundaries on the parameters such as magnetic field and electron density. A brief description of our theory appeared earlier in Ref. [\onlinecite{Joshua2013}], where it was used to interpret magneto-transport data, which were also presented in the same paper. In the present paper we provide more details of the theory and frame it in the broader context of a global phase diagram of the electronic states at the interface.

Our model includes itinerant electrons in the $t_{2g}$ bands of Ti that interact with local magnetic moments originating from electrons localized at the interface. The effective band structure we use for the three itinerant bands was discussed in Refs. [\onlinecite{Joshua2012,Zhong2013,Khalsa,Kim2013}] and is illustrated in Fig. \ref{fig:bands}(b). An important aspect of the band structure is that due to the confinement at the interface the $d_{xy}$ band is lower in energy than the $d_{xz}$ and $d_{yz}$ bands, so that at low densities only this band occupied. As the density of itinerant electrons is increased, the system undergoes a Lifshitz transition at a critical density, $n_c\sim 1.7\times 10 ^{13}\,\mrm{cm}^{-2}$, above which the the $d_{xz}$ and $d_{yz}$ bands are occupied as well~\cite{Joshua2012}.

The measured charge density of itinerant electrons at the interface is much lower than predicted by the polar-catastrophe model for formation of the conducting interface~\cite{Nakagawa2006}. The missing charges are thought to be sitting in localized states and possibly behave as local magnetic moments~\cite{Pentcheva2006}.
According to the polar catastrophe model the density of localized charges is expected to be
$n_{\mrm{imp}} \sim 3.3\times 10^{14}\,\mrm{cm}^{-2}$. Recent X-ray circular dichroism measurements~\cite{Lee2013} have located the magnetic moments on the Ti ions  and estimated the moment density to be $\sim 0.1\mu B/\mrm{Ti}$ (which corresponds to a density of $\sim 8\times 10^{13} \mrm{cm}^{-2}$), provided all the moments lie on a single atomic plane. Ab-initio studies suggest that only a fraction of the magnetic moments reside a few layers below the interface, preferentially near Oxygen vacancies, where they have a substantial interaction with the conduction electrons~\cite{Pavlenko2012}. Moreover, measurements using a scanning SQUID~\cite{Bert2011} indicate that many of the moments reside in dense clusters that are very far separated ($\sim 10\mu$) from each other and therefore do not affect the transport. Here we argue, that the large stretches of the apparently non magnetic region between these patches does contain a low concentration of local moments. We will consider the density of the uniformly distributed moments, which interact significantly with the conduction electrons, as a free phenomenological parameter.

A crucial new element in our theory is the interplay of two opposing magnetic interactions between the conduction electrons and the local moments. On the one hand, itinerant electrons in the $d_{xy}$ band interact with the local moments, for symmetry reasons, through antiferromagnetic exchange interactions. If the moments are not too dense this interaction leads to screening of the local moments through the Kondo effect~\cite{Brinkman2007a}. On the other hand the $d_{xz}$ and $d_{yz}$ bands couple to them through a ferromagnetic Hund's coupling, which opposes the tendency for screening. As we outline below and explain in detail in the rest of the paper, this competition between magnetic interactions leads to the global phase diagram shown in Fig. \ref{fig:pd:overview}. We also show how the transitions between these phases explain the distinct transport regimes seen in recent experiments.

We now give a brief overview of the paper and the main results.
The first part of the paper, sections \ref{sec:band} to \ref{sec:kondo} is devoted to a systematic derivation of the phase diagram shown in Fig. \ref{fig:pd:overview} from the minimal electronic model of the interface. The essential physics at zero field ($H=0$) can be understood as follows. At low electron density $n<n_c$ only the $d_{xy}$ band is occupied and the interaction of these electrons with the local moments is antiferromagnetic. The rough criterion for Kondo screening of the moments in this situation is $n>n_{\mrm{imp}}$ (for $n<n_{\mrm{imp}}$ we necessarily have under-screening), which gives the left phase boundary of the screened phase. At lower densities $n<n_{\mrm{imp}}$, RKKY interactions are expected to lead to Ferromagnetism~\cite{Michaeli2012}. Recent scans with magnetic force microscope indeed observed ferromagnetic domains at extremely low electron densities~\cite{Bi:arXiv:2013}.  Moreover, these measurements found a transition to a non magnetic phase when the density was increased using the gate voltage. We argue that this observation corresponds to the transition through the left boundary of the screened phase in Fig. \ref{fig:pd:overview}.

The Kondo screened phase established at $n>n_{\mrm{imp}}$, where the local moments are not extremely dilute,
may be viewed as a heavy-Fermi liquid in which the local electron moments are incorporated into the Fermi sea. The phenomenon is akin to the heavy fermi liquid established in metallic rare earth compounds, commonly understood within a mean field theory of a Kondo lattice model~\cite{Coleman:book}.

The  boundary of the Kondo screened phase at still higher electron density can also be understood quite simply. The
two higher bands, $d_{xz}$ and $d_{yz},$ begin to populate at the critical density $n_c$~\cite{Joshua2012}.  Electrons in these bands interact ferromagnetically with the local moments, competing with the antiferromagnetic interactions of the moments with $d_{xy}$ band. The competition leads to a monotonic reduction of the Kondo scale $T_K$ with $n$ for $n>n_c$. Hence, at some point the Kondo scale $T_K$ drops below the characteristic scale of magnetic RKKY interactions between the local moments. The critical density for this transition depends on the magnetic impurity concentration through the dependence of the RKKY scale on $n_{\mrm{imp}}$. The lower is $n_{\mrm{imp}}$ the lower the RKKY scale and hence the transition occurs at a higher critical density.
When then Kondo screening breaks down it gives way to magnetic phases driven by the RKKY interactions.  Most likely these are glassy phases in the limit of low impurity concentration, due to the random sign of the RKKY interaction.

When an in plane magnetic field is turned on in the Kondo screened phase, it will lead to substantial polarization of the local moments only when the associated Zeeman energy  exceeds the Kondo scale. This will happen in a meta-magnetic transition. The phase diagram in the $H$ versus $n$ plane will therefore also show the dome structure, which tracks the Kondo screening energy scale. We employ a large-N mean field theory as well as a perturbative RG, the latter valid in the limit of dilute magnetic impurities, to compute the Kondo scale and the critical magnetic field as a function of the mobile electron density. The variation of the critical magnetic field as a function of electron density within our model matches with the density dependent critical field at which a dramatic change is seen in the transport properties~\cite{Joshua2013}.
These experiments probed the high density side ($n>n_c$) of the phase diagram in the $H$ versus $n$ plane.

In Fig. \ref{fig:pd:overview} we  labeled some of the magnetic states, which appear outside of the Kondo screened phase at zero field. Such phases were discussed in several recent papers in relation to LAO/STO interfaces~\cite{Michaeli2012,Fidkowski2013,Banerjee2013,Pavlenko2013,Li2014}.
While this is not the focus of the present paper we would like to discuss them here briefly and explain how they fit in the context of the global phase diagram. First, when the local moment density is  high compared to that of the mobile electrons, RKKY interactions are ferromagnetic and are naturally expected to give rise to a Ferromagnetic state~\cite{Michaeli2012}. As the spin orbit coupling becomes appreciable at densities $n\sim n_c$ or more, the ferromagnetic order becomes unstable and yields to spin spiral states~\cite{Banerjee2013} or Skyrmion crystals~\cite{Li2014} with a rather long period. Finally when the localized moments are sufficiently dilute, on the right hand side of the paramagnetic lobe, the RKKY interaction is frustrated and may lead to glass order.

 We note that Pavlenko \textit{et. al.}~\cite{Pavlenko2013} predicted a different phase diagram consisting only of magnetic phases of the impurity spins. The difference stems from the fact that in Ref. \cite{Pavlenko2013} the impurities were assumed to be made of electrons localized in $e_g$ orbitals~\footnote{The model is based on earlier DFT calculations by the same authors~\cite{Pavlenko2012}.}. In this case the moments couple to the $t_{2g}$  conduction bands only through a Ferro-magnetic Hund's coupling, which cannot lead to Kondo screening. By contrast we assume that the important contributions to transport come from impurities localized to  $d_{xy}$ orbitals and therefore they couple through antiferromagnetic exchange to the itinerant $d_{xy}$ band.

The second part of the paper, section \ref{sec:transport}, investigates the transport properties implied by our model of the interface and by the phase diagram derived from it in the preceding sections.
We explain how the breakdown of Kondo screening can lead to the observed sharp drop in resistance with in plane magnetic field. In the limit of dilute magnetic impurities it is natural to understand the high resistance in the screened phase as a result of unitary scattering from the impurities. Polarization of the impurities with a field exceeding the Kondo scale reduces the scattering phase shift  making the impurities weak scatterers. We show that, in agreement with the model, the measured magnetoresistance curves that the system follows at different densities can be collapsed quite well to a universal function of $H/H_c$, where $H_c(n)$ is the measured critical field. One can also understand the sharp drop in resistance by considering breakdown of Kondo screening in a heavy Fermi liquid (HFL) phase where the screened impurity spins are incorporated into the Fermi sea. From this perspective the higher resistance of the screened phase stems from the small Fermi velocity in this phase.

Before proceeding we note that the Kondo effect has been observed on the surface of electrolyte gated STO~\cite{Lee2011}. However, since the transport in these samples is dominated by a single carrier type~\footnote{The Hall resistivity is almost perfectly linear at all densities unlike similar measurements in LAO/STO~\cite{Bell2009b,BenShalom2010,Joshua2012}}, the Kondo behavior fits the conventional picture where the Kondo scale rises continuously with the itinerant electron density. As we argue here, in LAO/STO where strong multi-carrier effects have been observed~\cite{Joshua2012} the Kondo scale is not expected to be monotonic in the density. We would also like to note that
the observation of the unique temperature dependence of the resistivity has been suggested very early on to be a demonstration of Kondo screening in this system~\cite{Brinkman2007a}.

\section{Three Band Model} \label{sec:band}
For the sake of completeness we briefly review the model of the conduction electrons
at the LAO/STO interface, which has been discussed in several recent papers~\cite{Joshua2012,Khalsa,Zhong2013,Kim2013}.

\begin{figure}
\centering
\includegraphics[width=8.5cm,height=5.5cm]{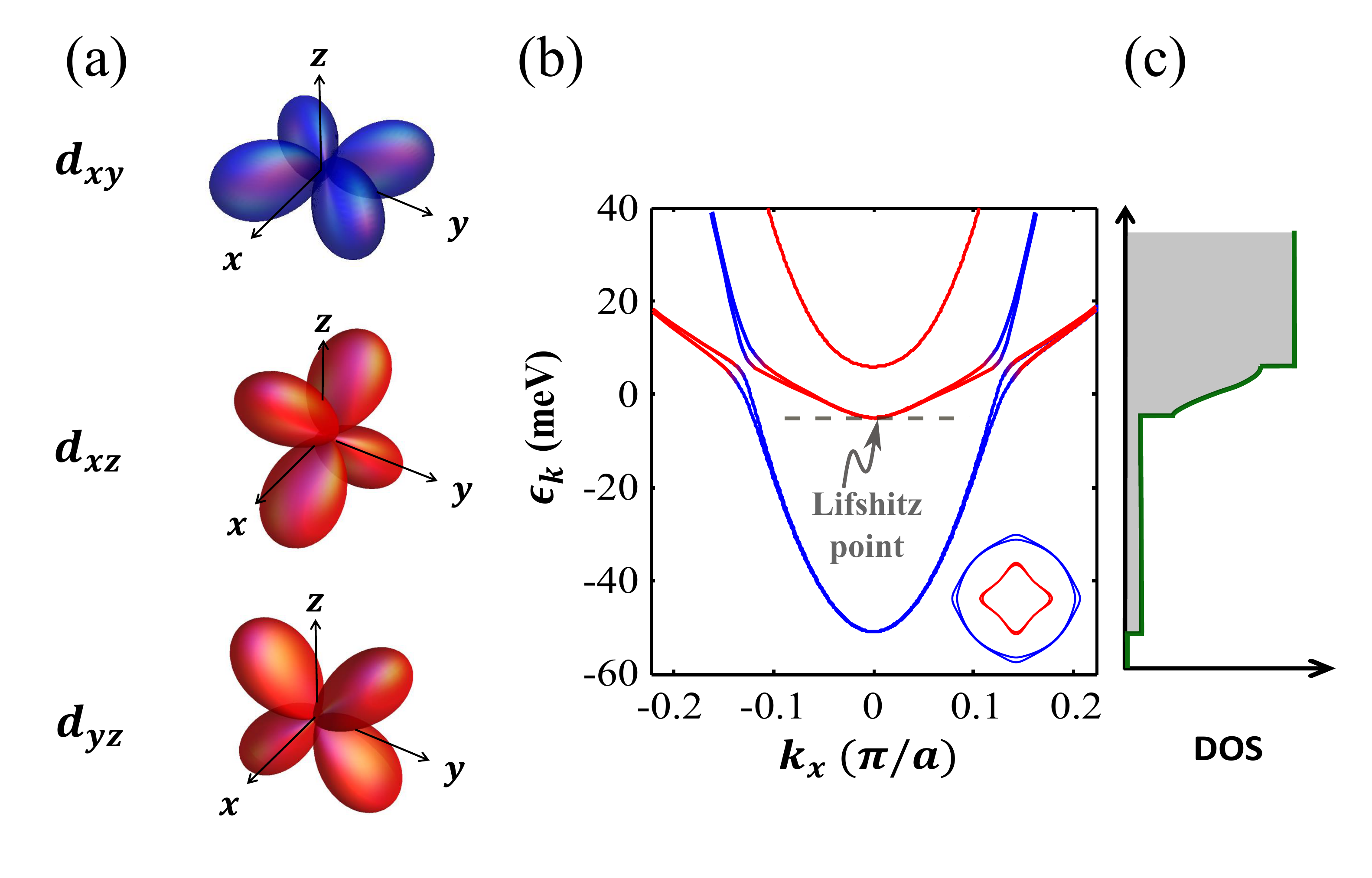}
\caption{(a) The spatial configuration of the three $t_{2g}$ orbitals of Ti ($d_{xy}$ in blue and $d_{xz}$,$d_{yz}$ in red). (b) The band structure of the Hamiltonian (\ref{H_c}) with $B=0$ with bands originating from the different $t_{2g}$ orbitals plotted using the color coding of panel (a). The spin splitting near the avoided crossing points is a result of the spin-orbit coupling, $\D_{SO}$, and the hybridization term, $\D_Z$.  The dashed line demarcates a Lifshitz transition point where new the two upper bands are first populated. The Fermi surfaces when the chemical potential is slightly above the Lifshitz transition (at $\mu=0$ on this scale) is presented in the inset. Panel (c) shows the density of states (arbitrary units) in this band structure as a function of energy. }\label{fig:bands}
\end{figure}

The model derives from the known structure of Bulk STO~\cite{Kahn1964} in which doped electrons go to the $t_{2g}$ orbitals of Ti (see Fig.\ref{fig:bands}.a). The confinement of the electrons along the $z$ direction at the interface lowers the energy of the $d_{xy}$ orbital because of its small hopping matrix element (large effective mass) along the $z$ axis. Besides the hopping matrix elements, the other ingredients of the model are: (i) the atomic spin-orbit coupling on the Titanium atoms (ii) the terms breaking the mirror symmetry at the interface and (iii) the coupling to an external magnetic field. Putting all of this together we can write the tight binding Hamiltonian as:
\be
{\mc H}_c=\sum_{\bs k} c\yd_{\bs k}\left( \hat H_{L} +\hat H_{SO}+\hat H_Z+\hat H_B \right)c\nd_{\bs k}\,,\label{H_c}
\ee
where the components of the electron creation operator $c\yd_{\bs k} = (c\yd_{\bs k \,1},c\yd_{\bs k\, 2},c\yd_{\bs k\, 3})$ correspond to the $d_{xy}$, $d_{xz}$ and $d_{yz}$ orbitals of Ti respectively.

Let us now specify the structure of the matrices $\hat H_L $, $\hat H_{SO}$, $\hat H_{z}$ and $\hat H_B$ starting with the lattice Hamiltonian $\hat H_L$ which describes the hopping matrix elements between adjacent $t_{2g}$ orbitals of Ti in the $xy$-plane~\footnote{We note that the hopping process involves an intermediate $p$-orbital state which belongs to the oxygen ion between each pair of Ti ions.
Therefore the hopping amplitudes should be viewed as effective ones obtained by integrating out these oxygen states.}. The lobes of the $d_{xy}$ orbital are arranged in the plane (see Fig.\ref{fig:bands}.a), such that it's hopping matrix elements in the $x$ and $y$ directions are equal and rather large $t=875 \, \mrm{meV}$~\cite{Mattheiss1972,Joshua2012,Santander-Syro2011,Zhong2013}. On the other hand, the lobes of the $d_{xz}$ ($d_{yz}$) orbital point out of the plane, giving rise to a smaller hopping element $t'\simeq 40 \, \mrm{meV}$ in the $y$ ($x$) direction and a large element $t$ in the $x$ ($y$) direction. Additionally, the $d_{xz}$ and $d_{yz}$ orbitals are hybridized by a diagonal hopping process with a hopping element $t_d \simeq t' $. Together these terms give the lattice Hamiltonian
 \begin{widetext}
\be
\hat H_{L} =
\begin{pmatrix} -t \left(\cos k_x+\cos k_y\right)-\D_E  & 0 & 0 \\
                 0 & -t \cos k_x -t' \cos k_y &  t_d \,\sin k_x \sin k_y \\
                 0 & t_d \, \sin k_x \sin{k_y} & -t' \cos k_x -t \cos k_y \end{pmatrix}
\,,\label{H_L}\ee
\end{widetext}
where $\D_E = 47\,\mrm{m\,eV}$~\cite{Santander-Syro2011,Joshua2012,Cancellieri2013} is the splitting induced by the confining potential along the $\hat z$ axis at the interface. All momenta here are rescaled by the lattice constant $a\approx 0.4 \,\mrm{nm}$.

The electric fields at the interface breaks the inversion symmetry there and thus allow for hybridization of the $d_{xy}$ orbital with the $d_{xz}$ and the $d_{yz}$ orbitals residing on neighboring sites along ${\hat{y}}$ and along ${\hat{x}}$. This leads to the  contribution
\begin{align}
\hat H_{Z} =\D_Z
\begin{pmatrix}  0    & i \sin k_y & i  \sin k_x \\
                    -i \sin k_y     & 0 & 0 \\
                    -i  \sin k_x     & 0 & 0 \end{pmatrix}\,,\label{H_Z}
\end{align}
Since the exact structure of the interface is not known we can only make a crude estimate $\D_z\approx 5 \, \mrm{meV}$). Note that to obtain this value one needs an interface field  $E_0 \approx 1 \,\mrm{V/nm}$ which is an order of magnitude larger than the breakdown field of $STO$ in the bulk. Hence our estimate of $\D_Z$ should be an upper bound to the real value. In any case the physics associated with this symmetry breaking term will not play an important role in the phenomena at the focus of this paper.

The atomic spin orbit coupling is given by
\begin{align}
\hat H_{SO} =&{\D_{SO}\over 2}\sum_{\a=x,y,z} \s^\a \otimes  L_\a \label{H_so}\\=& {\D_{SO}\over 2}
\begin{pmatrix} 0 & i\,\s^x &-i\, \s^y \\
                 -i\,\s^x & 0 &  i\s^z \\
                 i\, \s^y & -i\,\s^z & 0 \end{pmatrix}\,,\nn
\end{align}
Here $L_\a$'s are the angular momentum $l=2$ operators projected on to the $t_{2g}$ sub-space and $\s^\a$ are the Pauli matrices acting in the spin space. Since Ti is a rather light element the splitting $\D_{SO}=10\,\mrm{m\,eV}$ is small, and will be important only close to band degeneracy points.

The combination between the mirror symmetry breaking term (\ref{H_Z}) and the spin-orbit term (\ref{H_so}) gives rise to a Rashba like spin-orbit splitting~\cite{Petersen2000}. To see this let us focus on the $d_{xy}$ band near the $\G$-point (see the lower blue curve in Fig.\ref{fig:bands}.b near the dispersion minimum). Here the effective spin-orbit coupling is generated by the hybridization with the $d_{xz}/d_{yz}$ states (\ref{H_Z}) and (\ref{H_so}).
Since there is an energy gap $\D_E$ we may consider this effect in second order perturbation theory. Here the intermediate state will be virtual occupation of the $d_{xz}$ or $d_{yz}$ bands, such that the correction to the Bloch Hamiltonian of band $n=1$ is given by
\begin{align}
\d \hat H_1 = -\sum_{l=2,3}& {\langle \bs k,1|\hat H_Z |\bs k,l\rangle \langle \bs k,l|\hat H_{SO}|\bs k,1\rangle\over \e_{\bs k 1}^0 - \e_{\bs k l}^0}\nn\\ &\approx { \lam}\left[ k_y \s^x- k_x \s^y\right]\,,\nn
\end{align}
where $\lam ={ \D_{Z}\D_{SO} \over 2 \D_E}$ and where we have approximated the energy denominator by $\D_E$ and $\sin k_\a$ by $k_\a$.
Because of the small value of $H_Z$ and the large energy denominator the coupling close to the $\Gamma$ point in momentum space is very small, $\lam\approx 0.5\,\mrm{meV}$, negligible compared to the other effects.
The Rashba type coupling has a more noticeable effect near the degeneracy points between the $d_{xy}$ and $d_{xz}/d_{yz}$ bands where the energy denominator goes to zero and the perturbative calculation of $\lambda$ breaks down. This is demonstrated in Fig.\ref{fig:bands}.b where a splitting of all four spin and orbit states can be seen. Thus, the effects of the spin-orbit coupling are expected to peak near the Lifshitz point which has been observed in experiment~\cite{Caviglia2010a}.

Finally, the coupling of the conduction electrons spin and orbital moments to the external magnetic field is given by
\be
\hat H_B = -\mu_B \bs H\cdot \left(\bs L \otimes \mathds{1} +{g\over 2} \, \mathds{1}\otimes \bs{\s} \right)\,,
\ee
where $\mu_B$ is the Bohr magneton and $g$ is taken to be $ 2$.

\section{Magnetic couplings between conduction and localized electrons } \label{sec:moments}
The presence of magnetic moments is attributed to localization of charge near to the interface. In this region the localized states are expected to reside on the $d_{xy}$ orbital of Ti~\cite{Salluzzo2013,Lee2013}, which is lowered in energy by the confinement along $z$.
 A crucial point that has not been fully appreciated so far is that only a small subset of all the local moments, e.g. those bound to Oxygen vacancies at the right depth from the interface~\cite{Pavlenko2012}, may actually couple significantly to the mobile electrons forming the 2D gas. We will take this number of effective impurities to be a free parameter of the theory.

The minimal Hamiltonian of the system includes the contributions of (i) Hamiltonian of the itinerant electrons, (ii) the local Hamiltonian of the impurity sites and (iii) the hybridization between the two components. The three band model of the conduction electrons was discussed in the previous section. The Hamiltonian of electrons localized on the impurity site is given by
\begin{align}
\mc{H}_{\mrm{imp}} &=\sum_{i\in \mrm{imp}}\bigg[\sum_{l=1}^3\e_{ l} n_{i l} ^d + U \sum_{l} n_{i l\ua} ^d n_{il\da} ^d\label{H_d}\\&+U' \sum_{l\ne l'} n_{i l} ^d n_{il'} ^d -J_H \sum_{l\ne l'}\bs S_{il} \cdot \bs S_{il'}-\sum_l g \mu_B {\bf H}\cdot{\bs S_{il}}\bigg]\,,\nn
\end{align}
where $n_{il}^d = d_{il}^\dag d_{il}$ and $\bs S_{il} =  \half d_{il}^\dag \bs \s d_{il}$ are the density and spin operators of the local electrons on site $i$ and orbital $l = 1,2,3$. $\e_l$ is the single particle energy spectrum, where $\e_1<0$ binds a single $d_{xy}$ electron to each site $i$. $U$ and $U'$ are the intra and inter-orbital interaction strengths and $J_H$ is the Hund's rule Ferromagnetic exchange coupling between electrons residing on different orbitals.

\begin{figure}
\centering
\includegraphics[width=8cm,height=4.5cm]{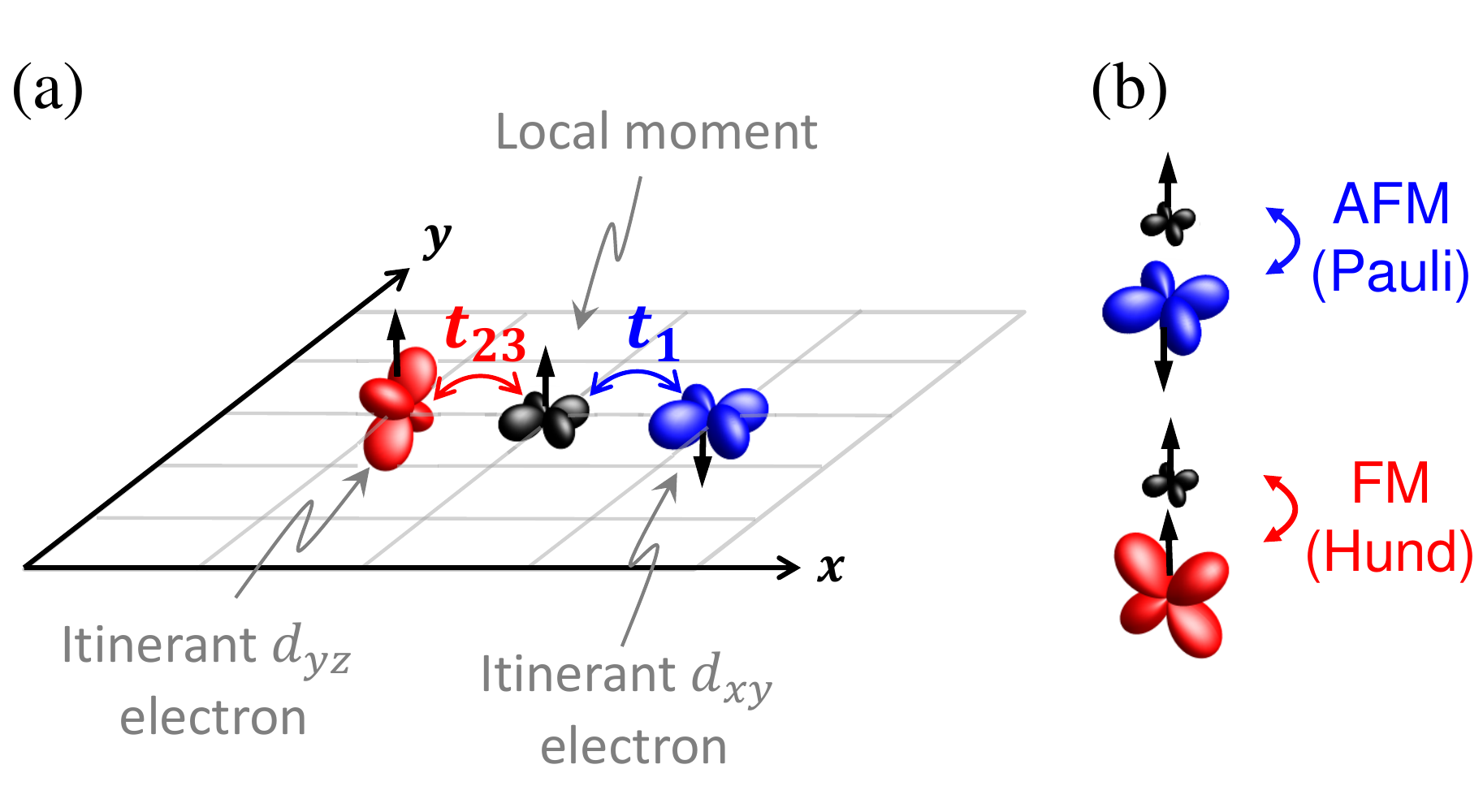}
\caption{Schematic representation of the virtual hopping processes between localized and itinerant electronic states. The local moments (black) are taken to be of $d_{xy}$ symmetry and the itinerant electrons can be of either $d_{xy}$ (blue), or $d_{xz}/d_{yz}$ (red) symmetry.  (a) Hopping between the $d_{xy}$ itinerant electron and a $d_{xy}$ localized state with amplitude $t_1$ and hopping between the itinerant $d_{xz}/d_{yz}$ electron and the unoccupied $d_{xz}/d_{yz}$ state on the moment site with amplitude $t_{23}$. Note that the local moment electron can also hop out of the impurity site to the $d_{xy}$ conduction band with amplitude $t_1$. (b) Processes of type (a) for the $d_{xy}$ result in an effective \textit{antiferromagnetic} coupling due to Pauli exclusion while the hopping of $d_{xz}/d_{yz}$ is allowed for both spin orientation and results in a \textit{ferromagnetic} coupling due to the Hund's rule coupling on the local moment site.}\label{fig:moments}
\end{figure}

The hybridization between the localized and itinerant electrons can be obtained using a tight-binding approximation (see Fig.\ref{fig:moments}). The largest hybridization occurs between orbital states of the same symmetry, both on the local moment and in the conduction bands. Therefore, to lowest order (up to nearest-neighbor in the tight-binding approximation) we have
\be
\mc{H}_{cd} =-\sum_{l=1}^{3}\sum_{\langle i,j\rangle}t_{l}^{ij}\left( c_{ j l}^\dag d_{i l}+h.c.\right)\,,\label{H_hyb}
\ee
where $i$ runs over the impurity sites and the brackets $\langle \rangle$ stand for summation over nearest neighbour sites in the two-dimensional plane. In this case $t_{2}^{ij} = t$ ($t_{2}^{ij} = t'$) if $i-j$ is a bond along the $\hat x$ ($\hat y$) direction, $t_{3}^{ij}$ has the same structure just rotated by 90 degrees and $t_{1}^{ij} = t$ is isotropic in all in-plane directions
Note that in the case in which the impurity is above or below the conducting layer $t^{ij}_l = t_l \d_{ij}$ and $t_{2} = t_{3}\gg t_{1}$ .
Putting all the terms together we have the total Hamiltonian
\be
\mc{H} = \mc{H}_c+\mc{H}_{\mrm{imp}} + \mc{H}_{cd} \,. \label{H_A}
\ee

 States with doubly occupied sites, i.e. either double occupation of a $d_{xy}$ orbital or occupation of one electron in the $d_{xy}$ orbital and one in a $d_{xz}$ or $d_{yz}$ orbital, are associated with large interaction energies of order $U$ and $U'$ respectively. We therefore integrate out the high energy charge fluctuations  on the local moment sites to second order in the hopping elements $t_1$ and $t_{23}$~\footnote{Note states with higher local occupation may be integrated out in the same manner.}. This way, using the standard Schrieffer-Wolff transformation we obtain effective magnetic interactions between the itinerant electrons and the local moments. Two types of processes contribute to the effective magnetic interactions (i) \textit{Particle processes} in which a conduction electron hops into a virtual state on the local moment site and back out with amplitude $t^2$ and an energy denominator of the order of $U$ or $U'$. (ii) \textit{Hole processes} in which the electron in the local site hops into the conduction band and back. Below we will explain how these processes give rise to an \textit{antiferromagnetic} coupling of the impurity spins to the itinerant electrons in the $d_{xy}$ band and {\it ferromagnetic} coupling of the impurity spins to the $d_{xz}/d_{yz}$ bands.

First consider the case where the intermediate state is in the $d_{xy}$ orbital. In this case both types of processes contribute. In the particle process an itinerant electron in the $d_{xy}$ band hops to the $d_{xy}$ orbital of the impurity site and then back, with the intermediate state energy $U+\epsilon_1$. Due to Pauli exclusion this process can occur only if the spin of the impurity site is anti-parallel to that of the itinerant electron (see Fig.\ref{fig:moments}.b). In the hole process the electron on the impurity site hops to the conduction band and back with the energy denominator being just $\e_1$ since we neglect the interactions in the conduction band. This process, of course cannot occur if the conduction band site is occupied with an electron with spin parallel to the impurity spin. Together these processes give rise to the antiferromagnetic exchange coupling
$J_1 \bs S_{i} \cdot \bs s_{i 1}\label{H_M1}$
with $J_1 = t_1 ^2[(U+\e_1)^{-1}-\e_1^{-1}]>0$ between the impurity spin ${\bf S}_i$ and the spin ${\bf s}_{i1}$ of the itinerant electron in the $d_{xy}$ band .

Now consider electrons in the $d_{xz}/d_{yz}$ orbitals. In this case the only contribution is from a particle process in which an itinerant electron from the $d_{xz}$/$d_{yz}$ bands hops on the $d_{xz}$/$d_{yz}$ of the impurity site and hops back. The Hund's rule coupling $J$ on the impurity site lowers the energy of the processes in which the impurity spin is parallel to the conduction electron spin Fig.\ref{fig:moments}.b. This leads to a ferromagnetic coupling $J_2 \bs S_i \cdot \bs s_{i l}$ with $J_2 = -t_{23} ^2 \left[ (U'+\e_2-J_H)^{-1}-(U'+\e_2+J_H)^{-1} \right]<0$, where we recall that $\epsilon_2$ is the single particle energy of the $d_{xz}$ and $d_{yz}$ states.

In summary after integrating out the charge fluctuations we obtain effective magnetic interactions
\be
\mc{H}_M = \sum_{i\in \text{imp}}\left( J_1 \bs S_i \cdot \bs s_{i1} + J_2\bs S_i \cdot\sum_{l=2}^3\bs s_{il} \right) \label{H_M}
\ee
with $J_1>0$ and $J_2<0$ and the sum is over impurity sites. Since charge fluctuations were integrated out, the effective Hamiltonian now acts in a constrained space with exactly one electron on each impurity site.

In writing $\mc{H}_M$,  we assumed for simplicity, that the impurity lattice lies directly above the conduction lattice, but this is of course not necessary. The impurity sites can just be random sites on the two dimensional lattice on which a localized state, and hence a local spin resides. Then the magnetic interactions will have the form ${\bf S}_i\cdot {\bf s}_{j(i) l}$, where $j(i)$ are the sites of the lattice neighboring the impurity site $i$.

The crucial point here is that the lower energy $d_{xy}$ band couples antiferromagnetically to the impurity spins whereas the higher energy bands  $d_{xz}$ and $d_{yz}$ couple to them ferromagnetically. Hence populating the two higher bands at the critical density also marks the   onset of a competing ferromagnetic exchange coupling to the impurity spins. In the next section we shall study the consequences of this competition.

\section{The Kondo screened phase and its breakdown in a magnetic field}\label{sec:kondo}
We now turn to investigate the phase diagram implied by the Kondo model in the space of itinerant electron density and magnetic field. The essential physics is the screening of the impurity spins at low temperatures. This screening is broken in presence of a magnetic field exceeding a critical value which depends on the electron density.  For the physical case of finite density of local moments we employ in section \ref{sec:mf} a mean field theory commonly used in Kondo lattice models of heavy fermion compounds~\cite{Coleman:book}.
We corroborate the results using an RG treatment of a single impurity  in section \ref{sec:rg},. This approach should be a good approximation in the limit of low impurity density.

\subsection{Mean-field approximation for the Kondo model}\label{sec:mf}

{We employ a mean field approximation for the system with many Kondo impurities in the spirit of the method commonly used to treat Kondo lattice systems \cite{Coleman:book}. This method usually relies on two levels of approximation: (i) the constraint of having exactly one electron on every local-moment site is relaxed and implemented only on average; (ii) a mean-field decoupling is applied to the Kondo interaction leading to effective hybridization between the local moment orbital and the conduction band.

Here we face the added complication, compared to the case of the Kondo lattice, that the local moments are disordered and therefore are not present on every conduction site.
 We solve this problem by treating the localized impurities as electrons, denoted by the second quantized operators $d\yd_{i\s}$ and $d_{i\s}$ that can potentially be present on {\em any site of the lattice}.
Thus these electrons form a flat band, with the average filling set to be $\sum_\s\av{d\yd_{i\s} d\nd_{i\s}}=n_{\mrm{imp}}$. In other words, we replace the localized spin-impurities by the same number of infinitely heavy electrons, interacting via the Kondo interaction with the conduction electrons. Of course, by mapping to a translationally invariant system we cannot capture the physics associated with incoherent scattering from impurities, but the energetics associated with opening of the Kondo gap is expected to be the same. In appendix \ref{app:disorder} we show how the same model can be derived from an alternative approach using quenched disorder averaging and assess the regimes in which the approximation is justified. }

The model of conduction bands coupled to a flat band representing the Kondo impurities is
\bea
\mc{H}_\lambda =\mc{H}_c+\mc{H}_M&-&\mu \sum_{\bs k}\sum_{l=1}^3 \left(c_{\bs kl}^\dag c_{\bs kl}-n\right)\nn\\
&-&\lam\sum_{\bs k}\left(d_{\bs k}^\dag d_{\bs k}-n_{\mrm{imp}}\right)\label{H_lam}.
\eea
Here $\mc{H}_c$ is the three band model of the itinerant electrons and the $H_M$ is the magnetic interaction hamiltonian of these electrons with the impurity electrons (\ref{H_M}), which is now extended to all sites of the lattice. The Lagrange multipliers $\mu$ and $\lambda$ are used to independently fix the average densities of the conduction band and impurity band respectively.

Now, that we are dealing with a Hamiltonian that acts in an unconstrained Hilbert space, we can apply the standard mean field approximation to decouple the magnetic interaction.
This is implemented through the variational hamiltonian
\begin{align}
\mc{H}_{MF}=& \mc{H}_c+\sum_{j=1}^N\left(\chi \,d_{j} ^\dag c_{j1}+h.c.  \right)\label{H_MF}\\-&\sum_{j=1}^N \sum_{l=1}^3 M_l s_{jl} ^x - \sum_{j=1}^{N} M_d S_j ^x\nn\\-&\mu \sum_{j=1}^N\sum_{l=1}^3 \left(c_{jl}^\dag c_{jl}-n_c\right)-\lam\sum_{j=1}^{N}\left(d_{j}^\dag d_{j}-n_{\mrm{imp}}\right),\nn
\end{align}
where we have decoupled the quartic interaction into the relevant channels involving the variational parameters $\chi$, $M_l$ and $M_d$. $\chi$ is a singlet hybridization field, which describes collective screening of the impurity moments. The parameters $M_l$ and $M_d$ account for the magnetization induced on the itinerant bands and the local moments respectively  by a field oriented along $S^x$. We  take $M_2=M_3$ to preserve orbital symmetry. Finally, the two Lagrange multipliers $\mu$ and $\lambda$ are used to fix the density in the conduction bands $n$ and the impurity band $n_{\mrm{imp}}$ independently.

\begin{figure}
\begin{center}
\includegraphics[width=8cm,height=13.5cm]{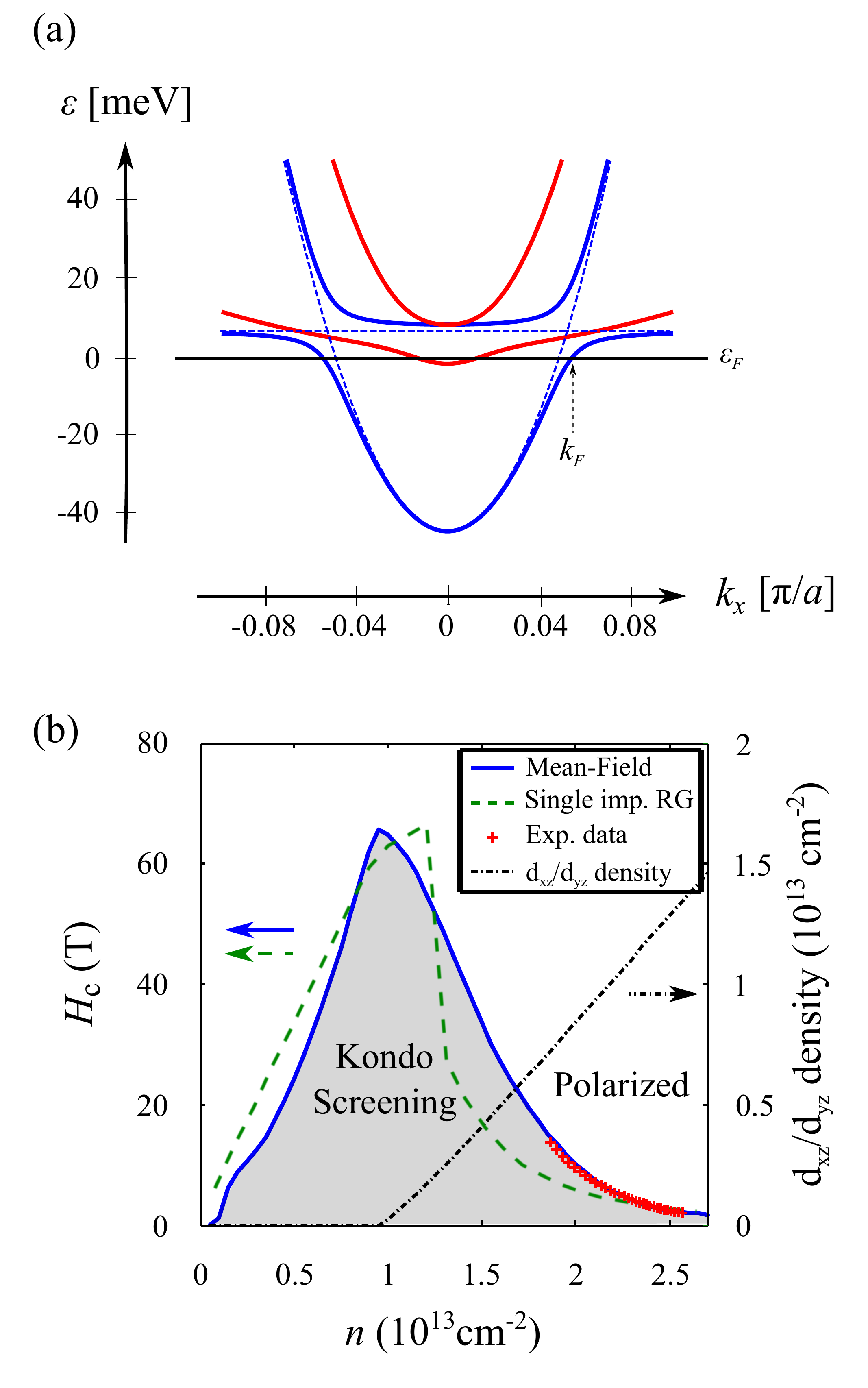}
\end{center}
\caption{(a) The band structure of the mean-field Hamiltonian (\ref{H_MF}) in the heavy fermi liquid phase. Here $\chi\ne0$ hybridizes the flat impurity band with the $d_{xy}$ conduction band leading to an expanded Fermi surface, marked by $k_F$ on the figure, which includes the total itinerant and local moment electron density.  The blue (red) curves correspond to $d_{xy}$ ($d_{xz}/d_{yz}$) bands, and the dashed line denotes the case with no hybridization field $\chi=0$. (b) Phase diagram for critical field vs itinerant electron density obtained by the mean field and perturbative RG calculations using the band-structure shown in (a). The blue solid curve is the first order transition line obtained from the mean field calculation for $J_1 = 0.9\,\mrm{eV}$ and $J_2 = -0.625\,\mrm{eV}$. It is compared to the critical field extracted from the experimental data in Ref.  [\onlinecite{Joshua2013}], shown in red (crosses).
The green dashed line gives an alternative evaluation of the characteristic field using RG calculation of a single magnetic impurity coupled to the multiple bands.  The critical field correspondds to the Kondo scale inferred from this calculation with  $J_1 = 0.55\,\mrm{eV}$ and $J_2 = -1\,\mrm{eV}$. The black dashed-dotted line gives the density of the $d_{xz}/d_{yz}$ conduction electrons, which grows continuously from zero at the Lifshitz point. The Kondo regime has a dome like shape peaked around the Lifshitz point.
 }\label{fig:pd}
 \end{figure}

Let us pause and consider the meaning of having a non vanishing hybridization field $\chi$. Such a term will mix the flat impurity band with the $d_{xy}$ band leading to the effective band structure illustrated in Fig. \ref{fig:pd}(a) (dashed and solid blue lines). The Fermi surface associated with the $d_{xy}$ band grows to encompass a Luttinger volume equal to the combined density of the itinerant $d_{xy}$ electrons as well as the impurity spins. Hence the Fermi wave vector will necessarily be at a point where the band curvature is affected by the hybridization with the flat band, leading to a
significantly reduced Fermi velocity. This is the essence of collective Kondo screening in systems with non-vanishing density of local moments and is the accepted explanation for the emergence of heavy electrons in rare earth metallic compounds~\cite{Coleman:book}.

To obtain the value of the variational parameters we minimize the expectation value  of the Hamiltonian $H_\lambda$ in the variational state generated as the ground state of $H_{MF}$. The minimum condition leads to the variational equations
\begin{align}
{\partial  \langle\mc{H}_\lambda \rangle \over \partial \left(\chi,M_l,M_d,\lam,\mu\right)} = 0 \label{Variational equations}
\end{align}
At zero field the solution is characterized by $\chi\ne 0$ and therefore describes a Kondo screened phase as explained above. When  the field is increased, a second minimum with $\chi=0$ forms, goes down in energy, and becomes the global minimum above a critical in-plane field. At this critical field, which marks a first order transition to the unscreened phase, the polarization of the local moments jumps from a small value to full polarization. The calculated value of the critical fields is plotted in Fig. \ref{fig:pd} as a function of the itinerant electron density.  To fit with the experimental data from Ref. [\onlinecite{Joshua2013}], plotted on the same figure, we used the parameters $J_1 = 900\,\mrm{meV}$, $J_2 = - 625\,\mrm{meV}$ and $n_{\mrm{imp}} = 0.26\times10^{13}\,\mrm{cm}^{-2}$. Note that the value of the magnetic couplings, especially the Ferromagnetic coupling $J_2$ are rather large.

It is important to note that the first order transition we obtain might be a spurious effect of the mean-field calculation. Moreover, any inhomogeneity due to disorder will in general revert the transition into a continuous percolation transition (according to Imry and Ma \cite{Imry1975}). The leading disorder effect is probably disordering of the local moments themselves. However, because the Kondo scale is strongly density-dependant the chemical potential disordering might also play an important role.

The Kondo screened phase forms a dome with a maximum critical field found at the Lifshitz density where the $d_{xz}$ and $d_{yz}$ bands begin to be populated. At $n>n_c$ the hybridization gap, and hence also the critical field for Kondo breakdown decreases due to the increasing competition from the ferromagnetic coupling with those bands.
On the other hand, the Kondo scale decreases as the density is lowered for $n<n_c$ as it must vanish when the electron density falls below the impurity density. The experiment in Ref. [\onlinecite{Joshua2013}] probed only part of the high density side of the dome.

The low density side of the dome, in zero field, may have been observed in recent magnetic force microscopy measurements reported in Ref. [\onlinecite{Bi:arXiv:2013}]. These experiments find that ferromagnetic domains suddenly appear only when the electron density is tuned to a very low value, well below the critical density $n_c$ associated with the Lifshitz transition.

 We note that the atomic spin orbit  coupling included in the above calculation is not essential for obtaining the dome shaped Kondo regime qualitatively. However without including the atomic SOC the decrease of $H_c$ on the high density side is extremely steep and cannot be well fit with the experimental result. The band mixing caused by the SOC smooths the density of states in the vicinity of the Lifshitz transition and thereby leads to a more gradual decrease of $H_c$, as seen in the experiment. Rashba-like SOC  generated due to the inversion symmetry breaking at the interface, could in principle, have an effect here too \cite{Isaev2012}, but in our case it would be too small to be noticeable.

{ Before proceeding we remark on the regime of validity of the above approach of treating the impurities as arising from a translationally invariant flat band. The end result of the analysis presented in appendix \ref{app:disorder}, indicates that the approach is valid in the limit where the Kondo screening clouds around an impurity spin is larger than the distance between impurities so that it encompasses many other impurities.
In the other, ultra-dilute limit, where the Kondo clouds are non-overlapping, then the correct approach is the single impurity RG discussed in the next section.}

\subsection{Perturbative RG for the multi-band system}\label{sec:rg}
We supplement the above mean-field result with a perturbative RG calculation
aimed at estimating the Kondo screening scale of a single impurity. This is done in the spirit of the poor man's scaling approach of Ref. [\onlinecite{Anderson1970}], generalized to three bands with competing ferromagnetic and antiferromagnetic couplings~\cite{Nozierrs1980,Zawadowski1980}.  Of course a single impurity will not undergo a phase transition but only a crossover as a function of the field from a screened state to a polarized state. The characteristic field of the crossover will be set by the Kondo scale.

The RG consists of rescaling the problem by successively integrating out states at the energy of the bandwidth $D$ perturbatively in the magnetic interactions. This leads to a flow of the dimensionless coupling constants $\Gamma_1=J_1 \nu_1$ and $\G_2=J_2\nu_2$ with the scale parameter $l=\log (D_0/D)$. Here the two couplings are
associated with the antiferromagnetic coupling of the impurity to the $d_{xy}$ electrons and the ferromagnetic coupling to the $d_{xz}$ and $d_{yz}$ electrons respectively. $\nu_{1,2}$ are the density of states of the respective bands at the cutoff scale $D$. The scaling equations of these coupling constants calculated to third order in the couplings are given by (see Refs. \cite{Nozierrs1980,Zawadowski1980} and appendix \ref{app:rg}  for details)
\begin{align}
{d\Gamma_1 \over d l} &=  2\Gamma_1 ^2-2\Gamma_1\left( \Gamma_1 ^2+\Gamma_2 ^2\right)+O\left(\Gamma^4\right)\nn\\ \label{RG equations} \\
{d\Gamma_2 \over d l} &=  2\Gamma_2 ^2-2\Gamma_2\left( \Gamma_1 ^2+\Gamma_2 ^2\right)+O\left(\Gamma^4\right)\,,\nn
\end{align}

A complication that we encounter in our case is that the different bands provide multiple cutoffs. The resolution of this problem is to separate the flow into two steps. First we integrate out only the deep hole excitations in the $d_{xy}$ band reducing the cutoff from its initial value $D_1=\mu+\D E$, i.e. the difference of the chemical potential from the bottom of the $d_{xy}$ band, down to $D_2=\mu$, the depth of the $d_{xz},d_{yz}$ bands. This gives the following scaling equations for the initial flow
\begin{align}
{d\Gamma_1 \over d l}& =  2\Gamma_1 ^2-2\Gamma_1^3 +O\left(\Gamma^4\right)\nn\\ \label{RG equations_mod} \,\\
{d\Gamma_2 \over d l}& = -2\Gamma_2\Gamma_1 ^2+O\left(\Gamma^4\right)\,,\nn
\end{align}
The full scaling equations  (\ref{RG equations}) are used in the second stage of the flow when
the running cutoff $D$ becomes lower than $D_2$.

Another complication is that the density of states of the bands is energy dependent. Of course this does not matter for the asymptotic flow at low energies but it is important for estimating a non universal number such as $T_K$. We leave the derivation of the full RG equations to appendix \ref{app:rg}.

\begin{figure}[ht]
\begin{center}
\includegraphics[width=5.5cm,height=5.5 cm]{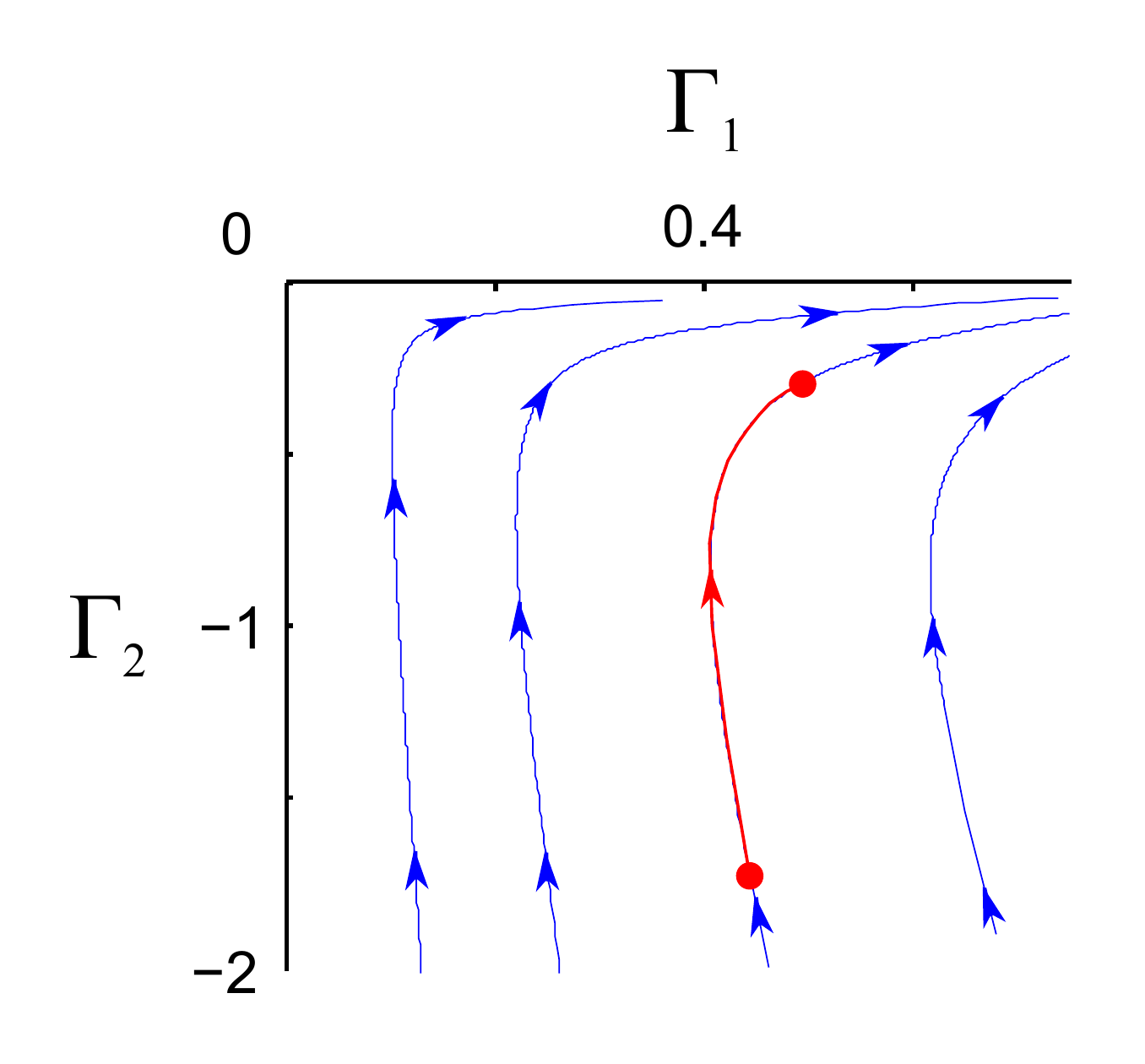}
\end{center}
\caption{The flow diagram of equations (\ref{RG1}) derived in appendix \ref{app:rg} for the case of $n = 1.3\times 10^{13}\,\mrm{cm}^{-2}$. The red line indicates the trajectory of the second stage of the flow for the values of $J_1$ and $J_2$ taken to produce the Kondo scale in the phase diagram Fig. \ref{fig:pd} ($J_1 = 550 \,\mrm{m}\,\mrm{eV}$ and $J_2 = -1 \,\mrm{eV}$).
 }\label{fig:flow}
\end{figure}

Let us now discuss the flow arising from the RG equations (\ref{RG equations}). Since these are perturbative equations obtained at third order in the couplings we should only consider the flow near the origin. At the outset of the flow we have $\G_1>0$ and $\G_2<0$, hence we focus on the bottom right quadrant of the flow diagram shown in Fig. \ref{fig:flow}. There, we see that the coupling $\G_2$ is irrelevant whereas $\G_1$ is relevant. We define the characteristic Kondo scale $T_K$ to be the value of the cutoff $D$ at which $\G_1$ reaches a value of $1/2$. Beyond that point the perturbative equations become unreliable but we expect the system to flow to the Kondo screened (local Fermi liquid) fixed point. The two fixed points at $(\G_1 = 1,\G_2 = 0)$ and $(\G_1 = 0,\G_2 = 1)$ are an unphysical artifact of the truncation of  (\ref{RG equations}) at third order in the coupling.

$T_K$ is highly sensitive to the initial value of $\G_2$ used in the second step of the RG flow. If it is large the band width $D$ needs to be reduced to a very small scale before $\G_1$ begins to grow. In this way the competition between the two couplings, $J_1$ and $J_2$ manifests itself in the RG flow.

The computed Kondo scale $T_K$ in units of Tesla is plotted in Fig.\ref{fig:pd} (the green dashed line) for $J_1 = 550 \,\mrm{m}\,\mrm{eV}$ and $J_2 = -1 \,\mrm{eV}$. At low densities, below the critical density $ 1\times 10^{13}cm^{-2}$ the Kondo scale $T_K$ grows linearly with density, as expected~\cite{Lee2011}. However, once the $d_{xz}/d_{yz}$ bands become populated $T_K$ first changes it's slope and then rapidly falls and becomes inversely proportional to the electron density $n_c$.
It is evident that the perturbative RG result does not fit the experiment as well as the mean-field analysis. This stems from the fact that the impurities have a finite density, which is taken into account in the mean field calculation while it is neglected in the single impurity RG scheme.

\section{Effect of the Kondo breakdown on the resistivity}\label{sec:transport}

Having derived the phase diagram, we now turn to discuss how the transition from a Kondo screened phase of the impurities to a polarized phase affects the observed transport properties of the system.
In particular we want to understand (i) the sharp drop of the resistivity observed as the in-plane field exceeds the critical value~\cite{BenShalom2010,Joshua2013} and (ii) the non linear, almost step-like dependence of the Hall resistvity as a function of magnetic field a a small tilt angle out of the plane~\cite{Joshua2013}.

We first explain the drop in resistivity in the limit of dilute magnetic moments, $n_{\mrm{imp}}\ll n$, which is indeed the relevant regime at conduction electron densities higher than the critical density  (right boundary of the screened phase in Fig. \ref{fig:pd:overview}). In this case  we consider the moments themselves as Kondo scatterers.
In the screened phase  the scattering phase shift of the single scatterer approaches the unitary limit leading to increased resistivity (This is the well known Kondo effect). In the case of a single band model the magneto resistance curves can be obtained exactly using the Bethe ansatz technique~\cite{Andrei1982}
\be
{\rho_K(x) } = \rho_0 \left({f_K(x) }+f_{\infty}\left[ 1-f_K(x)\right] \right)\,,\label{R_K}
\ee
where
\be
f_K(x)=\cos^2 \pi M(x) \,,
\ee
$x = H/H_K$ is the scaled magnetic field, $H_K$ is the Kondo scale translated to magnetic field units, $M$ is the magnetization of a single spin-half local moment. $\rho_{\infty} $ is the non universal saturation value of the resistivity. Here we are dealing with a more complicated three band model, so we expect to have a somewhat different scaling function of $H/H_K$.

\begin{figure}
\centering
\includegraphics[width=7.5cm,height=5.2cm]{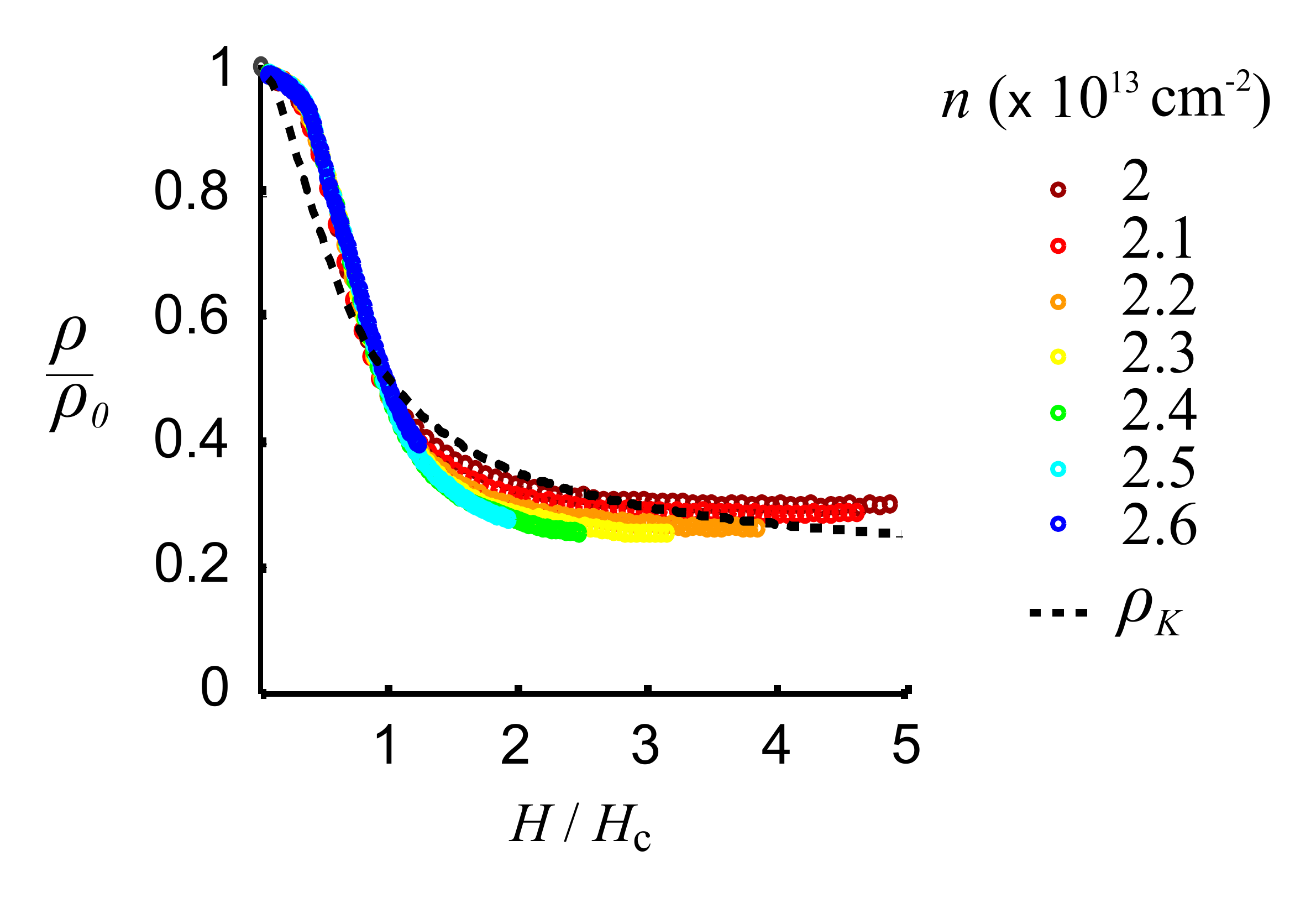}
\caption{  The measured normalized resistivity vs. the in-plane magnetic field $H$ scaled by the critical field $H_c$, which is plotted in Fig.\ref{fig:pd} (red crosses).  Here the different sets of data points represent different itinerant electron densities linearly distributed between 2 and 2.6 ($\times10^{13}\,\mrm{cm}^{-2}$). As can be seen, all curves collapse when scaled by $H_c$. The black dashed line is the universal MR curve obtained from the Bethe ansatz for a single Kondo impurity~\cite{Andrei1982} for comparison. }\label{fig:res}
\end{figure}

In Fig.\ref{fig:res} we present measured magneto resistance curves in parallel field at different electron densities. The data is extracted from the measurements described in Ref. [\onlinecite{Joshua2013}]. We see that upon scaling the $x$ axis by a characteristic field at each density the different curves collapse quite well on a common scaling function. Furthermore, these magnetoresistance curves are not far from the
Bethe Ansatz result (\ref{R_K}) with $H_K = H_c$ and $f_{\infty} = 0.2$, plotted for comparison with a (black) dashed line.

We note that reduction of conductivity on entering the Kondo screened phase may also receive a contribution from scattering on weak charge impurities in addition to the contribution from enhancement of scattering on the spin impurities. In the screened phase the flat impurity band is incorporated or hybridized into the Fermi, leading to reduction of the Fermi velocity. Breakdown of the Kondo screening at the critical field recovers the small Fermi surface of itinerant electrons with high Fermi velocity.

The conductivity due to scattering on {\em weak} charge impurities is given by $\sigma\sim k_F l= k_F v_F \tau$, where the relaxation rate in the Born approximation is $\tau^{-1}\sim n_i u^2 \rho(\epsilon_F) = n_i u^2 k_F/v_F$. Here $u\equiv \bra{k'}U\ket{k}$ is the matrix element for scattering between quasi-particle states on the Fermi surface and $n_i$ is the density of charge impurities. Hence we get
\be
\s\approx  \left({e^2\over 2\pi}\right)v_F^2 /(n_i u^2)
\ee
Since quasiparticles in the HFL have the same charge quantum number as the electron, scattering matrix elements low energy scattering due to weak charged impurities should be the same in the screened and polarized phases. In this case lower conductivity in the screened phase is due to the lower Fermi velocity.
Therefore, if we neglect the contribution of the $d_{xz}/d_{yz}$ bands to the conductivity, then  the ratio between the conductivities in the two phases is $\s^{HFL}/\s^P= (v_F^{HFL}/v_F^P)^2$. The mean field theory gives this ratio to be approximately $1/6$ in the range of densities above $n_c$, where the resistivity drop is seen in the experiments. However we note that such  a six fold drop of the resistivity is an upper bound on the effect of scattering on charge impurities. This is because the above
analysis is only true for weak scattering. In fact for very strong scattering the mean free path is simply set by the distance between impurities, which would not change with onset of Kondo screening. It is therefore possible that the entire drop of the resistivity is due to the spin scattering mechanism discussed above.

\begin{figure}
\centering
\includegraphics[width=8.9cm,height=4cm]{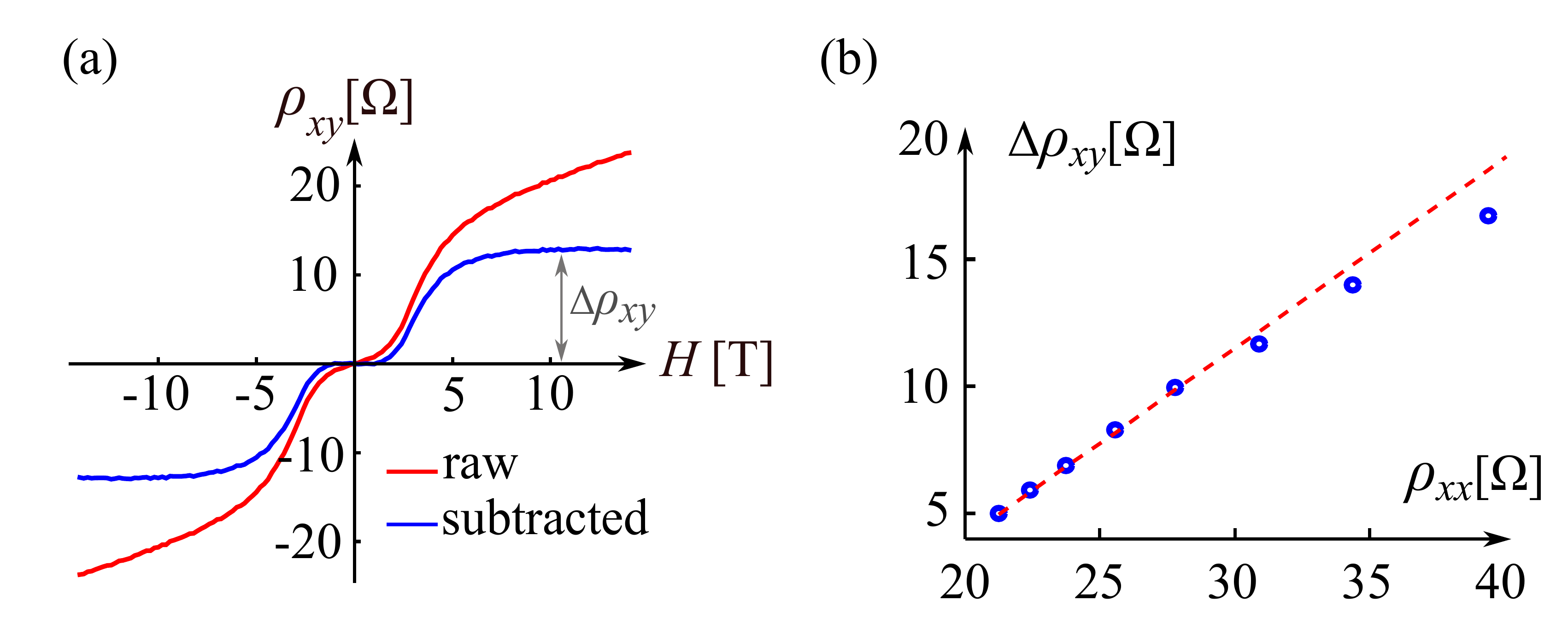}
\caption{ (a) Measurement of the anomalous contribution to the Hall effect. The red curve is the raw data. The blue curve is obtained by subtracting the ordinary linear Hall effect $\sim ( H/en) \sin \t$ where $\t=1^\circ$ is the angle between the plane and the tilted magnetic field. (b) The AHE contribution $\D\rho_{xy}$ extracted for different values of the gate voltage plotted against the longitudinal resistivity $\rho$ at $14\,$T where $\D\rho_{xy}$ is already saturated in most cases. The deviation from linearity observed at higher resistivity occurs in cases where $\D\rho_{xy}$ is not fully saturated at $14\,$T. The implied linear dependence of the anomalous contribution $\D\rho_{xy}$ on the scattering rate  $\propto \tau^{-1}$ suggests that the former stems from the mechanism of skew scattering~\cite{Nagaosa2010}.}\label{fig:AHE}
\end{figure}

Compelling evidence for the Kondo screening picture is provided by the step in the Hall resistivity as a function of almost in plane field, seen to occur together with the sharp drop in the longitudinal resistivity. We shall discuss how this effect can be understood in the context of the transition from a screened Kondo phase to polarized moments. The measurement conducted in Ref. [\onlinecite{Joshua2013}] was carried out in a field tilted by $\t=1^\circ$ degree out of the plane.
Fig.\ref{fig:AHE}(a) shows the evolution of the Hall resistivity as a function of the magnitude of the field. For $H<H_c$ the Hall resistivity exhibits the ordinary linear dependance on magnetic field, i.e. $\rho_{xy} = (H/en) \sin\t$. However, for $H>H_c$ the Hall resistivity rises dramatically over a small range of fields, until finally the effect saturates and $\rho_{xy}$ resumes the original linear dependence. Above the saturation field the Hall resistance has the form $\rho_{xy} = (H/en) \sin\t+\D\rho_{xy}$. This can be viewed as a strong indication for appearance of polarized magnetic moments above the critical field $H_c$. Once the moment is fully established the itinerant electrons are subject to a constant additional field induced by the polarized impurities through the magnetic exchange interaction with the conduction electrons, together with spin orbit interaction. Hence the observation of the anomalous component of the Hall resistivity gives further support for the Kondo screening picture whereby magnetic moments appear above a critical field.

We can gain more insight into the distribution of the magnetic moments from the characteristics of the AHE. In particular we can identify the specific mechanism by which they lead to anomalous Hall effect. For example in a system of dense moments the AHE is generated by the intrinsic mechanism due to a non vanishing Berry phase on the Fermi surface. In dilute impurity limit on the other hand, the anomalous Hall effect is generated by extrinsic mechanisms, that is in the process of scattering of electrons off the impurities in the same process that generates the resistivity. These extrinsic mechanisms are dominated by the process of skew scattering~\cite{Nagaosa2010,Nunner2008}.

To discern between these mechanisms we plot in Fig. \ref{fig:AHE}(b) the measured anomalous contribution to the Hall effect versus the resistivity.
The observed proportionality between the anomalous part of the Hall resistance $\D\rho_{xy}$ and the scattering rate is evidence that the extrinsic contributions are dominant. This in turn lends strong support to our estimate of the spin-impurity concentration as being much smaller than the density of itinerant electrons $n_{\mrm{imp}} / n \sim 0.1$. We note that other theories \cite{Pentcheva2006,Michaeli2012,Fidkowski2013,Banerjee2013} have assumed that the conduction electrons couple to a far larger density of impurity moments ($n_{\mrm{imp}} /n \sim 10$). In this limit
 the dominant contribution to the AHE should have been from the intrinsic mechanism.

\section{Conclusions}\label{sec:conclusions}

In this paper we presented a theoretical analysis of magnetic and magneto-transport phenomena observed at LAO/STO interfaces. We interpreted these phenomena as ariing from coupling between conduction electrons and spin impurities, which form due to localized electronic states at the interface. A competition between ferromagnetic and antiferromagnetic coupling of the impurities to the electrons in the different conduction bands leads to the magnetic phase diagram shown in Fig. \ref{fig:pd}.
the energy scale associated with Kondo screening of the impurities first grows with increasing gate voltage as the density of the $d_{xy}$ conduction electrons, which couple antiferromagnetically to the impurities, grows. Above the density $n_c$, the $d_{xz}$ and $d_{yz}$ bands begin to populate~\cite{Joshua2012} and the Kondo scale falls off due to the increasing competition from ferromagnetic coupling of the moments to these bands.

We have shown how the unconventional magneto-transport properties measured at densities above $n_c$ \cite{BenShalom2010,Seri2012,Joshua2013} can be understood in terms of the phase diagram shown in Fig. \ref{fig:pd}.b. First, the breaking of Kondo screening at the critical magnetic field leads to a sharp drop in the resistivity. In the limit of dilute magnetic impurities this can be understood in terms of a crossover from unitary scattering to weak scattering by the impurities. Another mechanism, effective at higher moment densities, involves the first order transition from a heavy Fermi liquid in the screened phase to a normal Fermi liquid (weakly renormalized) with a small Fermi surface in the polarized phase.
The Fermi velocity is increased when the screening breaks down, while low energy scattering by charge impurities remains unaffected leading to increased conductivity. Second, the breaking of Kondo screening and emergence of polarized moments above a critical field, together with spin-orbit scattering, gives rise to a large anomalous component of the Hall resistivity. This is seen in the experiments as a sharp rise, almost a step, in the Hall resistivity at the critical magnetic field when the field is tilted by a small angle out of the plane.

Interestingly, indications of the phase diagram of Fig. \ref{fig:pd} have also been seen in the low density regime below $n_c$. Measurements with a Magnetic force microscope revealed the presence of large ferromagnetic domains at very low densities~\cite{Bi:arXiv:2013}. The magnetic moments disappear, upon increasing gate voltage above a critical density as expected to occur when entering the screened phase from the left phase boundary in Fig. \ref{fig:pd}.

The excellent agreement between the theoretical model and the transport and magnetic measurements of the interface hold only if the density of the local moments is of the order of or smaller than the itinerant electron density, so that the Kondo screened phase is indeed established at a range of gate voltages. This implies that the number of local moments together with the number of mobile electrons
should be much less than the total charge that is expected to be transferred to the interface within the polar catastrophe model~\cite{Nakagawa2006}. There are however simple plausible explanations for this discrepancy. First, it is possible that much of the charge is found in doubly occupied localized states rather than in singly occupied "Anderson" impurities. Second, the local moments are likely to have a rather broad distribution in their distance from the interface. It is reasonable to expect therefore that only a small fraction of the moments couple significantly to the layer of conduction electrons.
Finally, we propose to investigate a more direct signature of Kondo screening by measuring the zero bias peak in the tunneling conductance through the LAO layer~\cite{Breitschaft2010}. In addition to the universal temperature dependance of the zero bias peak we also predict that the Kondo scale strongly depends on the density following the dome-like shape in Fig.\ref{fig:pd}.b.


\appendix

\section{Derivation of the effective magnetic couplings in a perturbative approach}\label{app:moments}
In this appendix we derive the effective magnetic couplings between the localized and itinerant electrons using the standard Schrieffer-Wolff transformation (SWT).
We start from the Anderson Hamiltonian (\ref{H_A})
\be
\mc H = \mc H_c + \mc H_{imp} + \mc H_{cd}
\ee
where $\mc H_c$ and $\mc H_{imp}$ are the conduction band and local moment Hamiltonians (\ref{H_c}) and (\ref{H_d}) respectively, and $\mc H_{cd}$ contains the hybridization terms between them. In the case that the itinerant electrons and the local moments lie in the same plane it is given by
\begin{align}
\mc H_{cd} =& \sum_{\langle ij \rangle} t c_{1j}^\dag d_{1i}+\rm{H.c.}\label{hyb_sup}\\
+&\sum_{i}  t c_{2i\pm \hat x}^\dag d_{2i}+t' c_{2i\pm \hat y}^\dag d_{2i}+\rm{H.c.}\nn\\
+&\sum_{i}  t' c_{3i\pm \hat x}^\dag d_{3i}+t c_{3i\pm \hat y}^\dag d_{3i}+\rm{H.c.}\nn
\end{align}
For low filling ($k_F a\ll 1$) his Hamiltonian can be simplified to a local hybridization term
\be
\mc H_{cd} \approx \sum_{i} t_l c_{li}^\dag d_{li}+\rm{H.c.} \label{Hcd:app}
\ee
since the momentum dependent Kondo interactions would be very small (of order $(k_F a)^2$).
Here $t_1 = 4t$ and $t_2 = t_3 = 2(t+t')$.

The SWT eliminates all states with non-unity occupancy ($n \ne 1$). The effective Hamiltonian in the singly occupied sub-space is computed to second order in the terms that couple the different occupancy sub-spaces:
\begin{align}
H_{\text{eff}}=\mc H_{11}+\mc H_{10}{1\over E-\mc H_{00}}\mc H_{01}+\mc H_{12}{1\over E-\mc H_{22}}\mc H_{21} \label{eH}
\end{align}
where we have neglected terms involving an occupancy higher than $n=2$. We note that the off-diagonal matrix elements $\mc H_{01}$ and $\mc H_{12}$ arise directly from the hybridization Hamiltonian (\ref{Hcd:app}).

The second and third terms on the left side of equation (\ref{eH}) give rise to the effective magnetic couplings. In the limit where $t_l \ll U,U',-\e_1$ they may be understood as second order processes in perturbation theory where the virtual states have zero and two electrons on the local moment site respectively. A crucial point is that both $\mc H_{22}$ and $\mc H_{00}$ , because they include a large interaction energy of order $U, U'$ or $-\epsilon_1$ respectively, have an average energy much larger than their bandwidth and much larger than the low energies $E$ of interest in the singly occupied subspace. Therefore we can replace the denominators by the charateristic gaps to the subspaces $0$ and $2$: $\mc H_{22} \simeq U$ or $U'$ (depending on the orbital configuration) and $\mc H_{00}\simeq -\e_1$.

Let us now write these terms explicitly taking into account these simplifications.
First let us consider the process in which the local moment site is empty in the virtual state, which contributes only to the coupling of the $d_{xy}$ conduction band
\begin{align}
 \mc H_{10}{1\over E-\mc H_{00}}\mc H_{01}\approx&-{{t_1 ^2 \over \e_1}}\sum_{i\s\s'} d_{i1\s}^\dag d_{i1\s'}c_{i1\s}c_{i1\s'}^\dag \mc{P}_0\,, \\& -{{t_1 ^2 \over \e_1}}\sum_{i} \left({1\over 2}c_{i1}^\dag c_{i1}+\bs \s_{i1}\cdot \bs  S_i \right)\,.
\end{align}
Here $\mc{P}_0 = (1-n_{i1-\s'})\prod_{l = 2}^3 (1-n_{il\ua}^d)(1-n_{il\da}^d)$ is a projector to the zero occupied state and $\bs S_i \equiv d_{i1}^\dag\bs \s d_{i1}$. Note that to obtain the last line we have used the identities $n_{1\ua}^d+n_{1\da}^d =  1$, $n_{1\ua}^d n_{1\da}^d = 0$ and $n_{l\s}^d = 0$ for $l\ne 1$.
Similarly, the doubly occupied processes give
\begin{align}
 \mc H_{12}{1\over E-\mc H_{22}}\mc H_{21}\approx -&{{t_1 ^2 \over U+\e_1}}\sum_{i} \left({1\over 2}c_{i1}^\dag c_{i1}-\bs \s_{i1}\cdot \bs  S_i \right)\nn\\
 &+\sum_{i,l=2,3} \left(K_2 c_{il}^\dag c_{il}-J_2 \bs \s_{il}\cdot \bs  S_i \right)
\end{align}
where $K_2 = {t_{2} ^2\over 2} \left[ {1\over U'+\e_2-J_H}+{1\over U'+\e_2+J_H} \right]$ and $J_2 = -t_{2} ^2\left[ {1\over U'+\e_2-J_H}-{1\over U'+\e_2+J_H} \right]$. Summing up these two contribution gives the Hamiltonian (\ref{H_M}), with $J_1 = {t_1^2 \over U}\left[{1\over U+\e_1}-{1\over \e_1} \right]$. Therefore, $J_1>0$ and $J_2 < 0$.

\section{Solution of the variational equations for the mean-field analysis}\label{app:mf}
In this appendix we elaborate on the solution of the variational equations (\ref{Variational equations}). Our goal is to compute the expectation value of (\ref{H_lam}) with the ground state wave-function of the mean-field Hamiltonian (\ref{H_MF}). To simplify we first make the two following approximations: (i) We expand the conduction band Hamiltonian (\ref{H_c}) around the $\G$-point, which is valid when $ka\ll \pi$. (ii) We neglect the terms that couple the $d_{xy}$ band to the $d_{xz}/d_{yz}$ bands, which greatly complicate the analytic expressions.
These spin-orbit terms are important only at high fillings and near certain $k$-points (namely, band crossing points), which do not have a particularly large DOS. Thus, taking these coupling into account will only slightly modify the results of the variational calculation.
Under these approximations the mean-field Hamiltonian (\ref{H_MF}) assumes a block-diagonal form
\begin{align}
\mc{H}_{MF} &=\label{H_MF_sup}\\& \sum_{k}\left[\begin{pmatrix} c_{k1} ^\dag & d_k ^\dag  \end{pmatrix}  \hat H_{1d} \begin{pmatrix} c_{k1} \\ d_k  \end{pmatrix}+\begin{pmatrix} c_{k2} ^\dag & c_{k3} ^\dag  \end{pmatrix}  \hat H_{23} \begin{pmatrix} c_{k2} \\ c_{k3}  \end{pmatrix} \right]\,, \nn
\end{align}
where
\begin{align}
&\hat H_{1d} =
\begin{pmatrix}
{k^2 \over 2m_l}-\tilde\mu-M_1 \s^x & \chi \\
\chi &-\lam-M_d\s^x
\end{pmatrix}\,, \label{H_1d} \\
&\hat H_{23} =\\&
\begin{pmatrix}
{k_x ^2 \over 2m_l}+{k_y ^2 \over 2m_h}-\mu-M_2 \s^x & -\D_dk_x k_y +i\D_{so} \s^z \\
-\D_dk_x k_y -i\D_{so} \s^z & {k_x ^2 \over 2m_h}+{k_y ^2 \over 2m_l}-\mu-M_2 \s^x
\end{pmatrix}\,,\nn
\end{align}
$\tilde\mu \equiv \mu +\D_E$ and $m_{l,h} \equiv 1/t_{l,h}a^2$ ($a \approx 0.39\,\mrm{nm}$).

The first step in the calculation is to diagonalize the Hamiltonian (\ref{H_MF_sup}) which gives the following quasi particle dispersions (see Fig.\ref{fig:pd}.a)
\be
\e_{k\s} ^{1d} = {1\over 2}\left({{k^2 \over 2 m_l}-\tilde\mu-\lam } -\s{M }\right) \pm A_{k\s} \label{HFL_dispersion}
\ee
and
\be
\e_{k\s} ^{23}  = \eta k^2-\mu\pm\sqrt{\D_{SO}^2+\left( W(\phi) \,k^2 +\s M_2   \right)^2 }
\ee
where $A_{k\s} \equiv \sqrt{\left({{k^2\over 2m_l}-\tilde\mu+\lam +\s \d M \over 2}\right)^2+\chi^2}$, $W(\phi) \equiv {1 \over \sqrt{2}} \sqrt{\d\eta ^2 +\D_d ^2 +(\d \eta ^2 -\D_d ^2) \cos 4\phi}$, $\eta = {1\over 2}\left({1\over 2 m _l }+{1\over 2m_h} \right)$, $\d \eta = {1\over 2}\left({1\over 2 m _l }-{1\over 2m_h} \right)$  and $\phi$ is the angle of $\bs k$.

Next, we use these dispersions and their corresponding quasi-particles to compute the expectation value of (\ref{H_lam}) which yields
\be
\langle\mc  H_{\lam} \rangle = \mc K - \mc N -\left(g\mu_B H\right) \mc S +\mc I_P + \mc I_K  \label{func}
\ee
where $\mc K$ is the average kinetic energy of the conduction electrons, $\mc N$ is the energy gain of particles due to the Lagange multipliers $\mu$ and $\lam$ and $\mc S$ is the total spin. The two terms denoted by $\mc I$ arise from the interaction term (\ref{H_M}). The first term, $\mc I_P = {J_1} \langle S \rangle \langle s_1  \rangle+J_2 \langle S \rangle \langle s_{23}  \rangle$, is the energy due to mutual spin polarization of the impurity band with the conduction bands, and the second term, $\mc I_K = -{J_1\over 2} \langle c_1^\dag d \rangle \langle d^\dag c_1 \rangle+h.c.$, is Kondo hybridization energy between band $l=1$ and the impurities. These two terms compete since the Kondo hybridization is a singlet state of the impurity electrons.

Finally, we wish to find the values of $M_1$, $M_2$, $M_d$, $\chi$, $\mu$ and $\lam$ which minimize the functional (\ref{func}), that is, to solve the set of equations (\ref{Variational equations}). We notice that $M_2$ may be determined immediately, namely $M_2 = -\mu_B H + J_2 \langle S \rangle $. Thus, we are left with five coupled equations. To solve them we take the derivatives of (\ref{func}) analytically and solve the equations numerically using a non-linear solver.

\section{Mean-field theory for the disordered Kondo model} \label{app:disorder}
{  Here we describe an alternative approach for deriving the mean field approximation described in section \ref{sec:mf} using quenched disorder averaging. For simplicity of presentation we consider here a Hamiltonian of a single conduction band coupled to the Kondo impurities:
\begin{align}
\mc H  = &\sum_{\bs k} (\e_{\bs k}-\mu) c_{\bs k }^\dag c_{\bs k }  \\
&+ {J\over 2} \sum_{j\in\mrm{imp}}  c^\dag_j {\bs \s} c_j \cdot{ \bs S}_j ,\nn
\end{align}
where $\e_{\bs k} = {k^2\over 2m}$ and the summation in the second term is over the impurity sites.

Now we make the mean-field approximation. As before, we will treat the impurity spins as a flat band of electrons and decouple the Kondo interaction to get an effective hybridization between the impurity
and the conduction band. But now we will not assume translational invariance. The lagrange multiplier $\lambda$ which was used in the main text to determine the average electron concentration in the impurity band is now taken to be site dependent. In this way it will act as a disorder potential, which can localize the moments on specific sites. This leads to the mean field Hamiltonian:
\begin{align}
\mc H_{MF}& = \sum_{\bs k} (\e_{\bs k}-\mu) c_{\bs k }^\dag c_{\bs k }\\
&+\chi_{\mrm{loc}} \sum_{j}\left(c_{j }^\dag d_j +\mrm{H.c.}\right)- \sum_{j=1}^N \lam_j\left(d_{j }^\dag d_j -n_j\right)\nn
\end{align}
Where  $\chi_{\mrm{loc}} = {J\over 2}  \sum_\s\langle c\yd _{i\s} d\nd_{i\s} +\mrm{H.c.}\rangle$. Below we will discuss what should be the variance of the disorder field $\lambda_j$.
In Fourier space this Hamiltonian takes the form
\begin{align}
\mc H_{MF} &= \sum_{\bs k} (\e_{\bs k}-\mu) c_{\bs k }^\dag c_{\bs k }\label{H_mf_app_4}\\
&+{\chi_{\mrm{loc}}} \sum_{\bs q \bs k}\rho_{\bs q}\left(c_{\bs{k+q} }^\dag d_{\bs k} +\mrm{H.c.}\right)
- \sum_{\bs q \bs k}\lam_{\bs q}\, d_{\bs{k+q} }^\dag d_{\bs k}\nn
\end{align}
where $\rho_{\bs q} \equiv {1\over N}\sum_{j \in \mrm{imp}}e^{i \bs q \cdot \bs R_{j}}$ is the density of impurity sites and $\lam_{\bs q} \equiv {1\over N} \sum_{j=1 }^N e^{i \bs q \cdot \bs R_{j}} \lam_j$. We separate the Hamiltonian into two parts: the translationally invariant part ($q=0$)
\begin{align}
\mc H_{0} =& \sum_{\bs k} (\e_{\bs k}-\mu) c_{\bs k }^\dag c_{\bs k 1}\label{H_mf_app_3}\\&+\chi  \sum_{ \bs k}\left(c_{\bs{k} }^\dag d_{\bs k} +\mrm{H.c.}\right)-\lam \sum_{\bs k} d^\dag _{\bs k} d_{\bs k }\nn
\end{align}
and the disorder potential ($q\ne 0$)
\be
\hat V = \sum_{\bs q\ne 0 ,\bs k}   \left( \chi_{\mrm{loc}}\,\rho_{\bs q}\,c_{\bs{k+q} }^\dag d_{\bs k} +\mrm{H.c.}-\lam_{\bs q} \,d_{\bs k+\bs q}^\dag d_{\bs k}   \right) \label{V}
\ee
where $\chi \equiv {n_{\mrm{imp}} }\;\chi_{\mrm{loc}}$ and $\lam \equiv {1\over N} \sum_{j = 1}^N \lam_j$.

The Hamiltonian (\ref{H_mf_app_3}) is identical to the flat-band model (\ref{H_1d}) with $M_1 = M_d = 0$. The  quasi-particle operators of the lower and upper bands respectively are given by
\begin{align}
&\psi_{\bs k-} = \cos \f_{\bs k} \,d_{\bs k}+\sin \f_{\bs k} \,c_{\bs k }\label{qp}\\
&\psi_{\bs k+} = -\sin \f_{\bs k} \,d_{\bs k}+\cos \f_{\bs k} \,c_{\bs k },
\end{align}
where $\cos \f_{\bs k} = {1\over \sqrt 2}\left[1+ {\e_{\bs k} - \mu +\lam \over \sqrt{\left(\e_{\bs k} - \mu +\lam \right)^2 +4\chi^2}} \right]^{1/2}$ and the corresponding dispersion is given by
\[\e_{\bs k} ^{\pm} = {1\over 2}\left[{k^2 \over 2m} -\mu -\lam  \pm \sqrt{\left({k^2 \over 2m} -\mu +\lam \right)^2+4\chi^2}\right],\]
In the weak coupling limit the values of $\chi$, $\lam$ and $\mu$, which are determined self-consistently, take the simple form
\begin{align}
&\mu \approx {n \over \nu}\label{mu}  \\
&\lam \approx -{\nu \chi^2 \over n_{\mrm{imp}}}\label{lam} \\
&\chi \approx {\sqrt{ n_{\mrm{imp}}\, n} \over \nu} \,e^{-{1\over 2J\nu}}\label{chi}
\end{align}
Here $n = \nu\,\e_F$ is the density of itinerant electrons and $\nu = m/2\pi$ is the density of states of the conduction band.

We see that the flat band model (\ref{H_1d}) emerges as the translationally invariant part of a disordered Hamiltonian. We shall now assess the effect of disorder. Specifically, we ask under what conditions the disorder does not significantly affect the Kondo gap $\chi$, which determines the global phase diagram of the model.

The disordered part of the hamiltonian (\ref{H_mf_app_4}), projected to the lower quasi-particle band, is given by
\be
\hat V_- = \sum_{\bs q\ne 0 ,\bs k} V_{\bs q}(\bs k) \psi_{\bs {k+q}- }^\dag \psi_{\bs k -} \,,\label{V}
\ee
with
\be
V_{\bs q} =  {\chi^2\over \lambda^2+ \chi^2}  { \left(\lam_{\bs q} + {\lambda \over  n_{\mrm{imp}} }\, \rho_{\bs q}\right)}
\label{VQ}
\ee
if $\bs k$ and $\bs{k+q}$ lie on the Fermi surface.
Within the lowest order Born approximation the decay rate of the quasi-particles due to the disorder potential (\ref{V}) is given by
\be
\Gamma_{\bs k} = {2\pi} \,\sum_{\bs q}\;\overline{V_{\bs q}V_{ -\bs q} }\;\delta(\e_{\bs k+\bs q}^{-} - \e_{\bs k}^{-}) \,.
\ee
To compute $\overline{V_{\bs q}V_{ -\bs q} }$ we need to know the strength of the disorder on the impurity band $\d \lam^2 =\overline{\lambda^2_j}-\lambda^2$. The latter can be estimated  to be  $\d\lam\sim \lam$.
Such disorder is large enough  to localize the $d$-electrons (impurities)  which gain a small dispersion of order $\lambda$ yet it is small enough so that the total number of the $d$ electron can still be effectively controlled by the average value $\lam$.
Thus $\overline{\lam_{\bs q} \lam_{-\bs q}}\sim \lam^2/N$.
For spatially uncorrelated disorder the variance of the impurity distribution $\rho_{\bs q}$ is independent of $\bs q$ and given by $\overline{\rho_{\bs q} \rho_{-\bs q}} = n_{\mrm{imp}}/N$. In this case the integration over the delta function gives the quasi-particle density of states ${1\over N} \sum_{\bs q} \delta(\e_{\bs k+\bs q}^{-} - \e_{\bs k}^{-}) = \nu \left( 1+{\chi^2 \over \lam^2}\right) $.

Because the impurity concentration $n_{\mrm{imp}}\ll 1$ the leading contribution to the  broadening is from the second term in (\ref{VQ}):
\begin{align}
\G={\chi^4\over (\chi^2+\lambda^2)^2}&\sum_q \left({\lambda\over n_{\mrm{imp}}}\right)^2\overline{\rho_{\bs q} \rho_{-\bs q}}\;\d(\e^- _{\bs k+\bs q} - \e^- _{\bs k})\nn\\&={\chi^4\over \lambda^2+\chi^2}\,{\nu\over n_{\mrm{imp}}}
\end{align}
Furthermore, because in the weak coupling limit $|\lambda|\ll \chi$, we have
 $\G \approx { \nu \chi^2 \over n_{\mrm{imp}}   } $.

 The decay rate $\Gamma$ should be compared to the Kondo gap $\chi$ in order to test how much it can change the phase diagram inferred from the variation of this gap. Using the weak coupling result (\ref{chi}) we find
 \be
 {\G\over \chi}={\nu\chi\over n_{\mrm{imp}}} = {l_{\mrm{imp}}\over \xi_K},
 \ee
where $l_{\mrm{imp}}=1/\sqrt{n_{\mrm{imp}}}$ is the distance between impurities and $\xi_K \approx \lambda_F \exp( {1\over 2 J\nu}) $ is the Kondo screening length of a single impurity.
This is a rather intuitive result. When the Kondo screening clouds of the different impurities are highly overlapping, each encompassing many other impurities, then they form a collective state which can be described as a heavy fermion band. On the other hand if the impurity concentration is so low that the screening clouds are essentially non overlapping then the single impurity physics applies.}

\section{Review of the two-channel Kondo RG equations}\label{app:rg}
In this appendix we review the derivation of the RG equations for the Kondo impurity coupled to multiple bands. We will follow the outline of the derivations in Refs. \cite{Hewson:book,Nozierrs1980} but with two added complications needed to obtain better quantitative estimate of the Kondo scale: (i) we allow the DOS to vary as a function of energy and (ii) allow the different conduction bands to have different band widths.

We begin by considering a single-channel Kondo problem but including an energy dependent density of states in the band. Our starting point is the Hamiltonian
\be
\mc{H} = \sum_{\bs k} \e_{\bs k}c_{\bs k }^\dag  c_{\bs k} +  {J} \sum_{\bs k \bs k'} \bs S \cdot c_{\bs k }^\dag\bs \s c_{\bs k'}\label{H_Kondo}
\ee
where $\e_{\bs k}$ is a general dispersion.

\begin{figure}[ht]
\begin{center}
\includegraphics[width=5.6cm,height=2.8 cm]{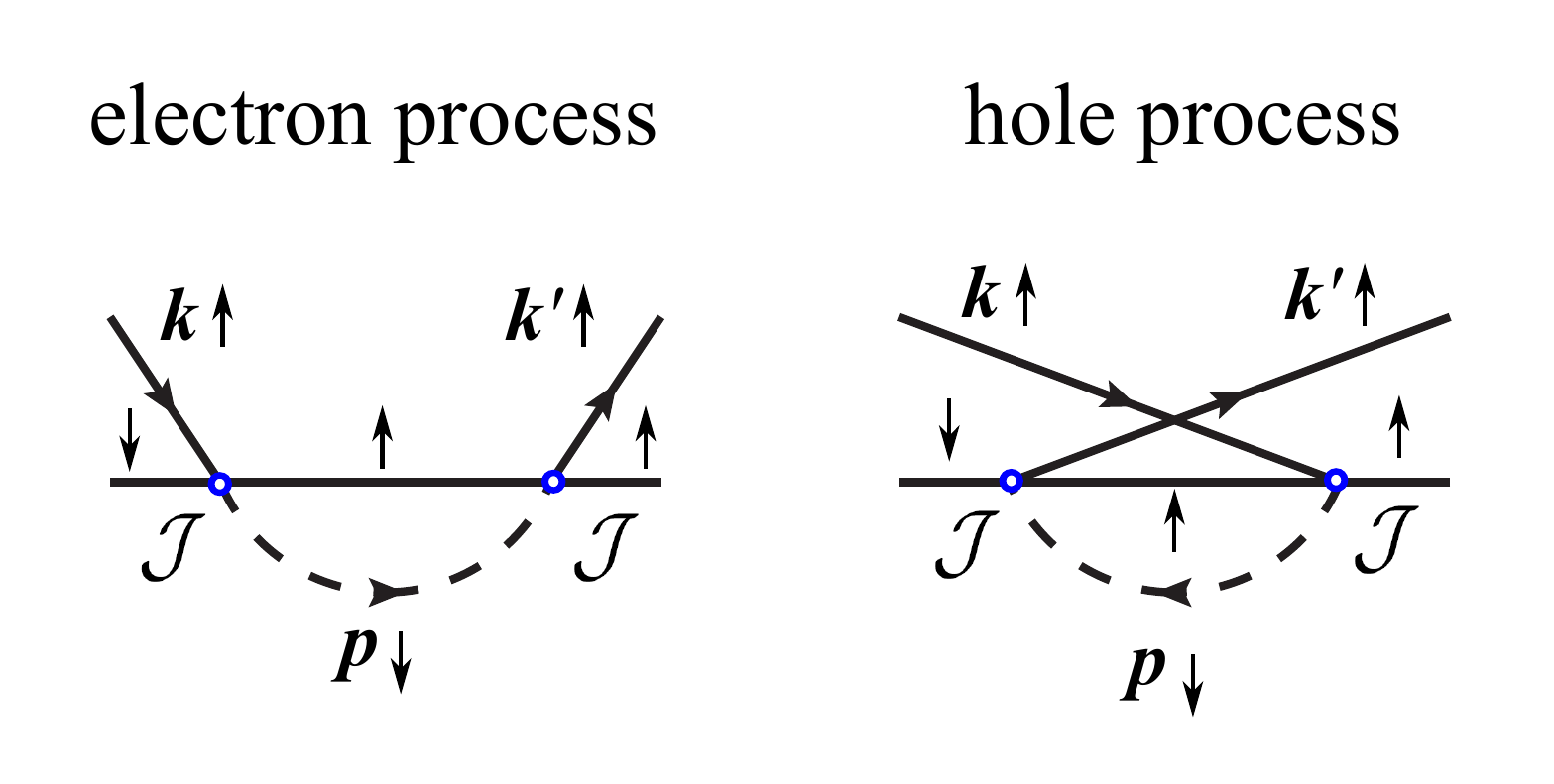}
\end{center}
\caption{Second order processes contributing to the renormalization of $ J$ which involve spin flips on both vertices. The dots represent a vertex of amplitude $ J$, the horizontal lines represents the local moment state and the solid and dashed lines represent conduction electrons close to Fermi level and from the high energy shell $D-|\d D|<\e_{\bs p}< D$ respectively. }\label{2nd order:fig}
 \end{figure}

To perform the RG transformation we integrate out a thin energy shell of width $|\d D|$ from the positive and negative band edges $\pm D$.
The second order processes be separated into two types, electron processes and hole processes (Fig.\ref{2nd order:fig}). In the case of an electron process an electron with initial energy $\e_{\bs k}\sim 0$ is scattered by the impurity to a virtual state in the energy shell $D-|\d D|<\e_{\bs p}< D$ and then scattered back to the Fermi level. For example,
such a process which includes a spin flip on both scattering events (left panel in Fig.\ref{2nd order:fig}) gives rise to the following correction term
\be
{ J^2} \sum_{p p' k k'}S^- \,c_{\bs k'\ua}^\dag c_{\bs p\da} {1\over E - D+\e_{\bs k}} S^+ \,c_{\bs p'\da}^\dag c_{\bs k\ua}\nn
\ee
Using the identity $S^- S^+  = 1/2 - S^z$ and $c_{\bs q} c_{\bs q'}^\dag = \d_{\bs q,\bs q'}$ this term can be brought to the form of (\ref{H_Kondo}).
Taking $\d D \rightarrow0$, $D\gg E,\e_{\bs k}$ and summing up all diagrams we get
\be
\d \mc{H}_{e} ^{(2)} \approx -{{J}^2\nu(D)} {\d D \over D} \sum_{\bs k \bs k'} \bs S \cdot c_{\bs k }^\dag\bs \s c_{\bs k'}\,,\nn
\ee
where $\nu(D)$ is the DOS at energy $D$. Similarly, for the hole processes we get the same result with $\nu(-D)$. Over all we have
\be
\d {J} ^{(2)} \approx -{2{J}^2}\overline\nu(D) {\d D \over D}
\ee
where $\overline{\nu}(D) = [\nu(D)+\nu(-D)]/2$.

\begin{figure}[ht]
\begin{center}
\includegraphics[width=7.7cm,height=5 cm]{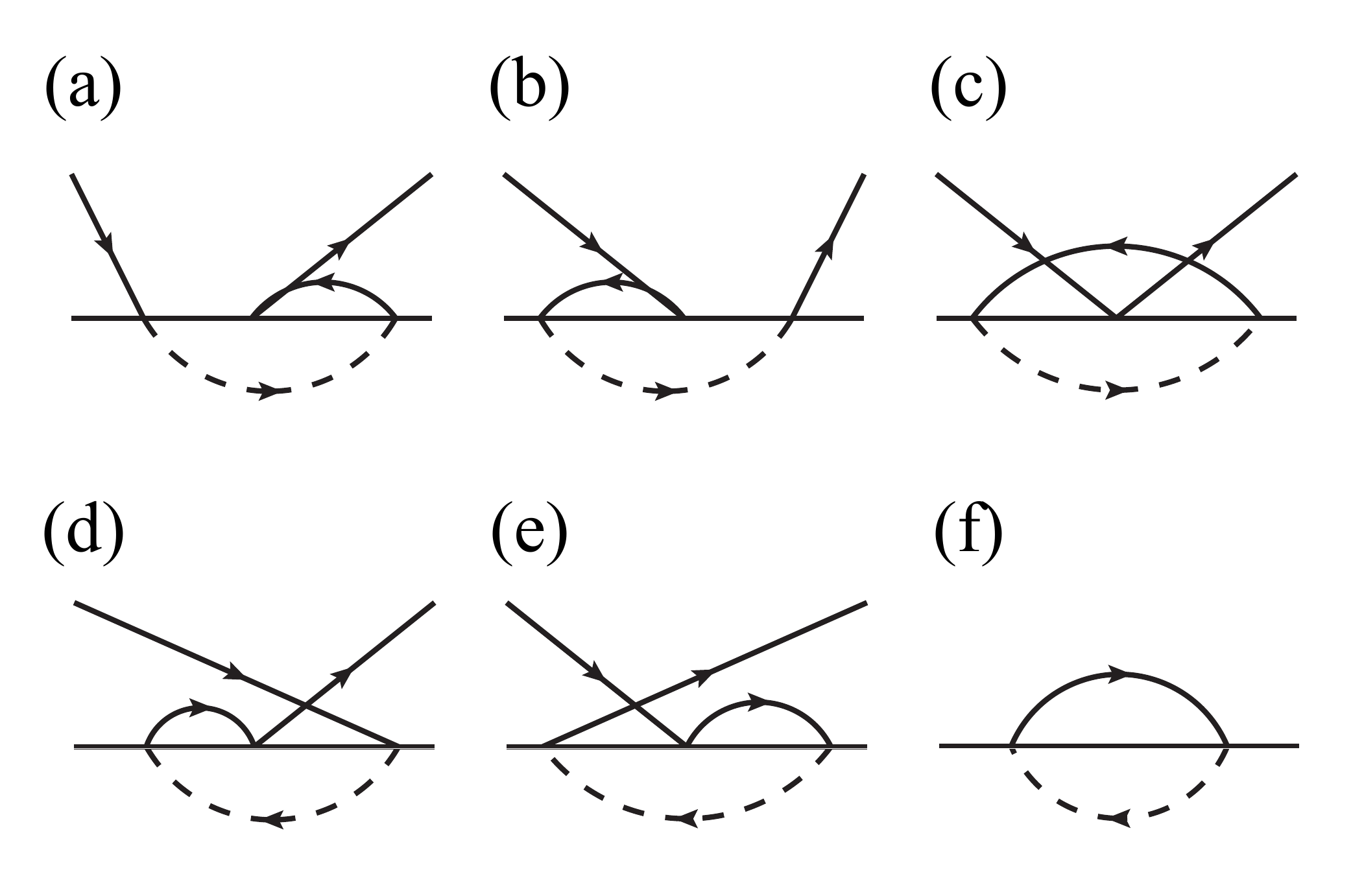}
\end{center}
\caption{}\label{3rd order:fig}
 \end{figure}

Let us now consider the third order diagrams contributing to the renormalization of $ J$. The diagramtic representation of the contributing processes to this order appear in Fig.\ref{3rd order:fig}.
In this case the diagrams include loops which correspond to integration over all momentum states from the Fermi energy down to the negative band edge. Such that the 3rd order correction has the form
\be
\d \mc{H} ^{(3)} \approx {{J}^3\overline \nu^2(D)} {\d D \over D} f(D) \sum_{\bs k \bs k'} \bs S \cdot c_{\bs k }^\dag\bs \s c_{\bs k'}\,,\nn
\ee
where
\be
f(D) =- {D\over \overline \nu(D)}\int_{-D} ^0 d\e \,{\nu(\e)  \over (\e-D)^2}\,.\nn
\ee
Additionally, since we are performing the RG transformation in the Hamiltonian formalism we must also account for the ground state energy shift due to the contraction of the diagrams presented in Fig.\ref{2nd order:fig} which is diagrammatically represented by Fig.\ref{3rd order:fig}.f
\begin{align}
\d E^{(2)}=&3{J}^2\overline \nu(D) \d D\int_{-D}^0 d\e\, {\nu(\e) \over E-D+\e } =\\
&3{J}\overline \nu(D)\d D \int_{-D}^0 d\e\, \nu(\e)\left({ {1\over D-\e}+{E\over (\e-D)^2}+... }\right)\nn
\end{align}
where $E$ is the energy of the effective Schr\"odinger equation
\be
\tilde H \psi = (E + \d E^{(2)})\psi \approx \left(1+{S}\right)E\psi
\ee
where $\tilde H$ is the renormalized Hamiltonian and
\be
S = 3{J}^2\overline \nu^2(D)  f(D) {\d D \over D}
\ee
Seeking the solution of the eigenvalue problem $|\tilde H  - (1+S)E|=0$ is equivalent to solving for the eigenvalues of $(1+S)^{-1/2} \tilde H (1+S)^{-1/2}$. Upon expanding, the effective Schr\"odinger equation assumes the form
\be
\left(1-{3\over 2} {J}^2\overline \nu^2(D)  f(D) {\d D \over D} \right)\tilde H \psi = E \psi
\ee
Keeping only terms to order of $O( J ^3)$ we have
\begin{align}
{d  J \over d\ln D} =  -\left[{  J^2\overline{ \nu}(D) } -  J^3 \overline{\nu}^2(D)f(D)\right]\label{RG_equation_sup}
\end{align}

We now turn to discuss the case of two channels. Each channel has an equation of the form (\ref{RG_equation_sup}) in addition to a coupling term. The coupling term arises from the diagrams Fig.\ref{3rd order:fig}.c where the middle vertex belongs to a different channel from the loop vertices. Additionally the energy shift of the ground state (see Fig.\ref{3rd order:fig}.f) contributes to both equations equally such that we have
\begin{align}
{d  J_1 \over d\ln D} =\nn&\\  - J_1&\left[{J_1\overline{ \nu}_1(D) } - J_1^2 \overline{\nu}_1^2(D)f_1(D)- J_2^2 \label{RG1} \overline{\nu}_2^2(D)f_2(D)\right]\nn\\ \\
{d  J_2 \over d\ln D} =\nn& \\ -J_2&\left[{  J_2\overline{ \nu}_2(D) } - J_2^2 \overline{\nu}_2^2(D)f_2(D)- J_1^2 \overline{\nu}_1^2(D)f_1(D)\right]\nn
\end{align}
where $f_{1,2}$ and $\overline \nu_{1,2}$ are defined with the DOS of band 1 and 2 respectively.
To account for the asymmetric bandwidths $D_1\ne D_2$ we simply take $\nu_\a(\e)=0$ for $\e>D_\a$.

\end{document}